\documentclass[fleqn,usenatbib]{mnras}

\usepackage{newtxtext,newtxmath}

\usepackage[T1]{fontenc}

\DeclareRobustCommand{\VAN}[3]{#2}
\let\VANthebibliography\thebibliography
\def\thebibliography{\DeclareRobustCommand{\VAN}[3]{##3}\VANthebibliography}

\usepackage{graphicx}	
\usepackage{amsmath}	

\usepackage{float} 

\usepackage{array} 
\usepackage{upgreek} 

\title[How is cold gas affected by magnetic fields?]{How is cold, star-forming gas in galaxies affected by magnetic fields?}

\author[Kamran R. J. Bogue et al.]{
Kamran~R.\ J.\ Bogue,$^{1}$\thanks{E-mail: kammybogue@gmail.com}
Rowan~J.\ Smith,$^{2}$\thanks{E-mail: rjs22@st-andrews.ac.uk}
Robin~G.\ Tre{\ss},$^{3,4}$
Mordecai-Mark Mac Low,$^{5}$ \newauthor
David~J.\ Whitworth,$^{6,10}$
Ralf~S.\ Klessen,$^{3,7,8,9}$
Noé~Brucy,$^{3,10}$
Philipp Girichidis,$^{3}$
Simon~C.~O.\ Glover,$^{3}$
\newauthor
Junia~Göller,$^{3}$
Juan~D.~Soler,$^{11}$
and
Alessio~Traficante$^{11}$ 
\\
$^{1}$Jodrell Bank Centre for Astrophysics, Department of Physics and Astronomy, University of Manchester, Oxford Road, Manchester M13 9PL, UK\\
$^{2}$School of Physics and Astronomy, University of St.\ Andrews, St.\ Andrews, Scotland, UK\\
$^{3}$Universit{\"a}t Heidelberg, Zentrum fur Astronomie, Institut f{\"u}r Theoretische Astrophysik, Albert-Ueberle-Str. 2, 69120 Heidelberg, Germany\\
$^4$Laboratory for Galaxy Evolution and Spectral Modeling, EPFL, CH-1015 Lausanne, Switzerland \\
$^{5}$Department of Astrophysics, American Museum of Natural History, 200 Central Park West, New York, NY 10024, USA\\
$^{6}$Universidad Nacional Aut\'onoma de M\'exico, Instituto de Radioastronom\'ia y Astrof\'isica, Antigua Carretera a P\'atzcuaro 8701, Ex-Hda., Mexico\\
$^{7}$Universit\"{a}t Heidelberg, Interdisziplin\"{a}res Zentrum f\"{u}r Wissenschaftliches Rechnen, Im Neuenheimer Feld 225, 69120 Heidelberg, Germany\\
$^{8}$Harvard-Smithsonian Center for Astrophysics, 60 Garden Street, Cambridge, MA 02138, USA\\
$^{9}$Elizabeth S. and Richard M. Cashin Fellow at the Radcliffe Institute for Advanced Studies at Harvard University, 10 Garden Street, Cambridge, MA 02138, USA\\
$^{10}$ENS de Lyon, CRAL UMR5574, Universite Claude Bernard Lyon 1, CNRS, Lyon 69007, France\\
$^{11}$INAF-IAPS, Via Fosso del Cavaliere, 100, 00133, Rome, Italy
}

\date{Accepted XXX. Received YYY; in original form ZZZ}

\pubyear{2024}

\begin{document}
\label{firstpage}
\pagerange{\pageref{firstpage}--\pageref{lastpage}}
\maketitle

\begin{abstract} 
Numerical simulations provide a unique opportunity to improve our understanding of the role of magnetic fields in the interstellar medium of galaxies and in star formation. However, many existing galaxy-scale numerical simulations impose a Kennicutt-Schmidt (KS) star formation law by construction. In this paper, we present two {\sc{Arepo}} simulations of an isolated star-forming galaxy with and without magnetic fields, using sink particles to model star formation without imposing a KS relation. We examine global differences between the models, and investigate the impacts on star formation. We include a time-dependent, non-equilibrium chemical network coupled to a thermal evolution scheme and supernova feedback. Our magnetic field amplifies via dynamo action from a small initial seed field. We find a more compact magnetohydrodynamic (MHD) disc (radius $\sim$ 5.1\,kpc, compared to $\sim$ 7.4\,kpc), with a diffuse atomic envelope above and below the plane that is not seen in the hydrodynamic (HD) case. The HD disc displays a smoother, more even radial distribution of gas and star formation, and more bubbly substructure. Our MHD simulation has a higher proportion of dense, gravitationally unbound gas than the HD case, but a lower star formation rate, an average between 125--150\,Myr of $\sim 4.8\,\mathrm{M_{\odot}}$~yr$^{-1}$ compared to $\sim 8.4\,\mathrm{M_{\odot}}$~yr$^{-1}$. We see a clear shift in the KS relation to higher gas surface densities in the MHD case, more consistent with observations. The additional magnetic support against gravitational collapse seems to raise the threshold gas surface density required for star formation.


\end{abstract}

\begin{keywords}
galaxies: star formation -- galaxies: ISM -- ISM: magnetic fields -- methods: numerical -- MHD
\end{keywords}



\section{Introduction}
\label{sec:intro} 
 
Magnetic fields have been a persistent challenge to gaining a complete picture of star formation in galaxies. It has long been understood that the presence of interstellar magnetic fields will provide an additional magnetic pressure that can prevent gravitational collapse \citep{Mestel1956StarClouds}. Even so, quantifying the precise impact of magnetic fields on the formation of stars from cold, dense gas in galaxies remains an active area of research \citep{Hennebelle2019TheEvolution}. The star formation process is complex, multi-scale, and occurs in the highly dynamic environment of the interstellar medium \citep[ISM, see also][]{Draine2010PhysicsMedium, Klessen2016PhysicalMedium}. Improving our understanding of the ISM and the forces that drive its evolution is crucial for studying star formation. In this work, we address the question of how magnetic fields affect the ISM within galaxies, and thus what effects they have on galactic-scale star formation.

Due to the paucity of accurate measurements of magnetic field morphology and strength, a complete picture of the ISM has historically been elusive. However, significant progress has been made in the last few decades in the observation of magnetic fields \citep[see Section 2 of][for a review on magnetic field measurements]{Pattle2022MagneticCores}, which has developed our understanding of their effect on the ISM. It has become clear that magnetic fields thread the ISM on all scales \citep{Han2017ObservingFields}, and data from nearby galaxies show that magnetic energies are sufficient to be dynamically important in the ISM \citep{Beck1996GALACTICPerspectives}. \citet{Crutcher2012MagneticClouds} infer from Zeeman observations that most molecular clouds have magnetic fields strong enough to compete with turbulence, and that the field can be dominant \citep[see also][]{Heiles2005TheTurbulence, Whitworth2025OnAnalysis}. All-sky polarisation maps from the \textit{Planck} satellite have provided an unprecedented insight into the magnetic field morphology of our own Milky Way galaxy \citep{PlanckCollaboration2016PlanckResults}. The correlation between gas structures and magnetic field morphology seen in the \textit{Planck} maps, and in cloud-scale observations from the Balloon-borne Large Aperture Submillimeter Telescope for Polarimetry \citep[BLASTPol;][]{Galitzki2014TheFlight} and the Submillimeter Polarimeter for Antarctic Remote Observations \citep[SPARO;][]{Li2006ResultsClouds}, provide evidence that on size scales greater than 1\,pc, the energetic importance of magnetic fields is typically equal to or larger than that of turbulent gas motions \citep{Pattle2019SubmillimeterRegions}. This suggests magnetic fields play an important role in the formation of molecular clouds, and thus star formation.

Numerical simulations have become a powerful avenue for expanding our understanding of the importance of magnetic fields in the ISM. In (500\,pc)$^3$ stratified boxes with periodic boundary conditions,  \cite{Girichidis2018TheClouds} find that magnetic fields slow fragmentation, and that gas fragments into fewer clumps than in simulations without a magnetic field. They additionally find the galactic disc in their magnetohydrodynamic (MHD) simulation is thicker by a factor of a few than the disc in their purely hydrodynamical (HD) simulation. A key result from MHD simulations is the significant suppression of star formation  \citep{Federrath2013ONCLOUDS, Kortgen2015ImpactEvolution}. Using stratified box simulations, \citet{Brucy2023Large-scaleGalaxies} find that the star formation rate (SFR) can be suppressed with increasing field strength, with an increase of $10\,\upmu\mathrm{G}$ reducing the SFR by a factor of 10. For theoretical models with magnetic field strengths comparable to observed values, the SFR seems to be suppressed by a factor of 2--3 compared to the value in unmagnetised models \citep[see][and references therein]{Krumholz2019TheFunction}. 

Critically, however, the suppression of star formation depends strongly on whether or not the magnetic energy is dominant in comparison to thermal motions, turbulence, and gravity. \citet{Seta2025MagneticFields} argued using Zeeman measurements and pulsar dispersion measures that magnetic fields are in equipartition with the turbulent kinetic energy at all densities in the ISM, while \citet{Heiles2007MagneticClouds}
used H\,{\sc i} Zeeman observations to argue that the fields are dynamically dominant in the diffuse ISM, though this is not the regime in which star formation occurs. McGuiness et al. (in prep) suggest that magnetic energy dominates the energy budget across the thermal instability regime of the ISM, from $n \sim 1 \rm \,cm^{-3}$ to densities where gravitational collapse becomes dominant in the simulations, $n \sim 10^4 \,\rm cm^{-3}$, in agreement with earlier results from \citet{Seifried2020SILCC-Zoom:Fields,Ibanez-Mejia2022GravityClouds}.

Many of the studies discussed above do not capture the large-scale galactic environment. However, large-scale galactic dynamics has profound effects on the gas that evolves within the galaxy \citep{Smith2020}. Spiral arms and bars can trigger the formation of dense clouds from warm gas \citep{Tress2020SimulationsGalaxy}, and the global rotation curve determines the local stability and shear experienced by the ISM \citep{Li2005StarCollapse, Meidt2018AMotions, Meidt2020ADecoupling}. Simulations of galaxy discs with resolved molecular clouds must simulate a broad dynamic range of scales to recover all of these effects. 

Galactic magnetic fields result from the amplification of small seed fields by both the small-scale or turbulent dynamo \citep{Kazantsev1968} and the large-scale or mean-field dynamo \citep{Brandenburg2005AstrophysicalTheory}. The small-scale dynamo is driven by turbulence. It amplifies the field exponentially with a time constant of the local eddy turnover time, typically a few  megayears in the ISM, producing fields with spatial scales under the eddy size of $100 \, \rm pc$ \citep{Korpi-Lagg2024ComputationalGalaxies}. In modern star-forming galaxies, supernovae (SNe) represent the primary source of turbulent driving \citep[][]{Balsara2005AmplificationChaos, Steinwandel2020OnGalaxies}, although radial transport becomes important for more gas-rich galaxies \citep{Krumholz2016IsTurbulence}. For further detail on the effect of the small-scale dynamo, we refer the reader to other works that show how the small-scale dynamo amplifies magnetic fields in the ISM \citep[e.g., see ][]{Sur2010THESTARS, Schleicher2010Small-scaleLimit, Federrath2011ATURBULENCE, Schober2012MagneticNumbers, Rieder2016APhase, Rieder2017AConfiguration,Gent2021Small-scaleTurbulence}. The large-scale dynamo, which is driven by differential rotation, amplifies the field over much larger timescales and spatial scales, approximately the rotation time of the galaxy disc, of a few hundred megayears.

Numerical studies are now beginning to be able to simulate galaxies with a broad range of spatial scales, whilst also including the effects of magnetic fields. In dwarf galaxy simulations, \cite{Whitworth2022MagneticGalaxies} showed that in their low metallicity models, magnetic fields did not suppress the overall SFR, though they did change the molecular gas content. However, increasing in mass up to the scale of the Milky Way, most simulations do see a suppression of SFR \citep{Kim2020LocalSpurs/Feathers, Wibking2023TheGalaxies}. These results are consistent with the result of \citet{Pakmor2013SimulationsGalaxies} that the equilibrium dynamo field strength depends on the mass of the galaxy, with equipartition only being reached for galaxies with mass $M > 10^{10}\,\mathrm{M_{\odot}}$. Milky Way-scale models have also suggested that magnetic fields limit the growth of feedback bubbles \citep{Robinson2024RegulatingGalaxy, Zhao2024FilamentaryFormation}.

It is clear that magnetic fields at realistic field strengths can have significant impacts, affecting disc fragmentation, the formation of giant molecular clouds, and suppressing SFRs \citep{Kortgen2018TheGalaxies, Kortgen2019GlobalGalaxies}. These studies also allow us to determine what field strengths are necessary to produce such effects. Concerning SFRs for example, \cite{Su2017FeedbackFormation} showed that magnetic field strengths of $10^{-2}\,\upmu\mathrm{G}$ have little effect, whereas fields of order $10\,\upmu\mathrm{G}$ as in \cite{Wibking2023TheGalaxies} can suppress SFRs by factors of 1.5--2. See also \citet{Dobbs20232aScales} for an in-depth discussion. 




The Kennicutt-Schmidt \citep[KS;][]{Schmidt1959TheFormation.,Kennicutt1989TheDisks} relation links the gas content of galaxies to their SFRs and is key to understanding galaxy evolution. It relates the SFR surface density to that of the total neutral gas (H\,\textsc{i} + H$_2$) via $\Sigma_{\rm SFR} \propto \Sigma_{\rm gas}^{n}$, with an exponent of $n\sim$\,1.4. Observations also show a linear correlation between the SFR surface density of galaxies and the molecular hydrogen surface density---the molecular KS relationship \citep{Bigiel2008TheScales}. As observations improve, the KS relation has been studied for a wide range of objects, and at different scales. Recent work, compiling results that included lower mass objects such as dwarf galaxies, has confirmed the original result \citep{Reyes2019RevisitingGalaxies, Kennicutt2021RevisitingLaw} of an exponent $n \sim 1.4$ for the combined sample, though dwarf galaxies yield a slope closer to $n\sim 1$ \citep{Filho2016THEGALAXIES, Roychowdhury2017ExtendedGalaxies}. The effect of magnetic fields on this relationship is actively being studied \citep{Whitworth2022MagneticGalaxies, Brucy2023Large-scaleGalaxies}.



We approach the question of how magnetic fields affect star formation numerically. We perform two simulations of isolated galaxy discs using {\sc{Arepo}}, an MHD and an HD case. Galactic-scale simulations often impose either a KS law, or a set star formation efficiency per free-fall time within their prescription for star formation \citep[e.g.][]{Wibking2023TheGalaxies, Robinson2024RegulatingGalaxy}. In this work, we utilise sink particles, which only form and accrete when gas is identified to be unambiguously gravitationally bound, as determined by the dynamics of the gas. In this way, we do not force a KS relation--it emerges purely as an output from the simulations. Early examples of the use of sink particles in hydrodynamic galaxy scale simulations include \citet{Li2005StarCollapse, Li2006StarCollapse}. In our MHD model, as we also allow our magnetic field to emerge as an output, this allows us for the first time to measure a self-consistently generated KS relation across an entire galactic disc for both an HD and MHD simulation. Both simulations in this work are run from the same initial condition and with the same custom physics modules. This allows us to interrogate the effect of the field specifically. Furthermore, the MHD model is run from an initial seed field to give a field structure self-consistently generated by a dynamo. In Section~\ref{sec:methods} we describe our implementation of the {\sc{Arepo}} code, and the initial conditions of our models. We present the results in Section~\ref{sec:results} and in Section~\ref{sec:discussion} discuss the implications of our findings and important caveats. We conclude in Section~\ref{sec:conclusions}. 



\section{Methods}
\label{sec:methods} 

\begin{figure}
    \centering
    \includegraphics[width=0.45\textwidth]{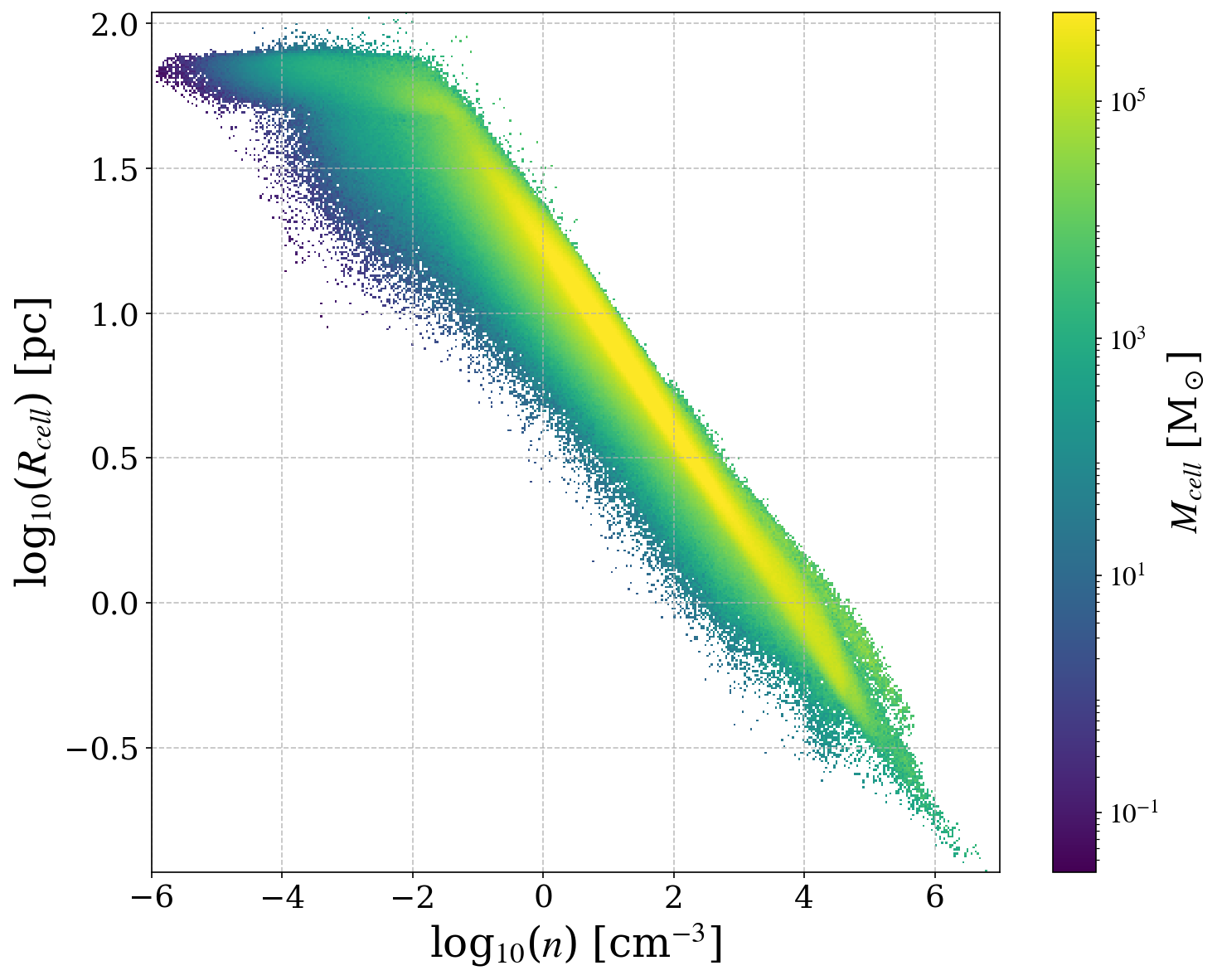}
    \caption{Variation in cell size with gas density for the MHD model.}
    \label{fig:resolution}
\end{figure}

The {\sc{Arepo}} code has often been used to simulate isolated galaxies, and is well described by \citet{Springel2010}, \citet{Pakmor2013SimulationsGalaxies}, and \cite{Pakmor2016ImprovingAREPO}. Here we use a custom version of {\sc{Arepo}}, that includes modules designed to capture the physics of the cold ISM and star formation, following a method similar to \cite{Tress2020SimulationsGalaxy}. This version was first used with the inclusion of magnetic fields in dwarf galaxy models by \cite{Whitworth2022MagneticGalaxies}, for more recent work see also \citet{kjellgren2025dynamicalimpactcosmicrays}. We note here the most important features for the reader to understand our findings.

\subsection{Numerics}
\label{subsec:code} 

Our modified version of {\sc{Arepo}} includes custom physics modules to model magnetic fields and star formation in the cold ISM. {\sc{Arepo}} solves the ideal MHD equations in three dimensions, as described by \cite{Pakmor2011MagnetohydrodynamicsGrid} and \cite{Pakmor2013SimulationsGalaxies}. On galactic scales, non-ideal MHD effects such as ambipolar diffusion are unimportant \citep{Wurster2020Non-idealFormationb, Wurster2020Non-idealFormation}. 

The MHD equations are solved by {\sc{Arepo}} on an unstructured mesh constructed at each time step from the Voronoi tessellation of a set of mesh-generating points. These mesh-generating points are moved with the flow of the simulated fluid to enable a pseudo-Lagrangian method, which enables good numerical resolution in high-density regions. At interfaces between cells, {\sc{Arepo}} computes the flux by solving the Riemann problem in the rest-frame of the interface, employing up to three solvers depending on the result. The Harten-Lax-van Leer Discontinuities (HLLD) Riemann solver \citep{Miyoshi2005AMagnetohydrodynamics} is first, and if this returns a valid result then no other solver is used. However, in extreme circumstances, this can return negative values for density and pressure. If so, the solution is not valid and is discarded. In this case, the more-diffusive HLL solver \citep{Harten1983OnLaws} is tried next. If this still returns a negative solution, then the solution is discarded once again, and the Rusanov solver \citep{Rusanov1962TheObstacles} is employed, which guarantees a valid positive solution, but is strongly diffusive. This process is used in both the MHD and HD simulations discussed later, to avoid numerical differences between the two models. 

Discretisation errors that arise in solving the MHD equations cause a non-zero divergence of the magnetic field. This is corrected by the \citet{Powell1999AMagnetohydrodynamics} divergence cleaning scheme. This divergence problem is discussed in \cite{Pakmor2013SimulationsGalaxies}, which first described the implementation of the Powell approach in {\sc{Arepo}}. An analysis showing that this approach is valid for galaxy-scale simulations using {\sc Arepo} is presented in the appendix of \cite{Tress2024MagneticDynamics}.

We use the default mesh refinement scheme, in which the code refines and derefines locally, as necessary, in order to keep the mass of each mesh cell within a factor of two of a target mass. In the models presented in this work, the target mass is set to 1000\,M$_{\odot}$. As the density of gas increases, the cell size decreases, and, particularly important in cold gas, the Jeans length also decreases. Generally, to avoid artificial fragmentation, the Jeans length must be resolved by at least 4 cells \citep{Truelove1997TheHydrodynamics, Federrath2011ATURBULENCE}. In order to satisfy this condition we set the size of our sink particles to 5\,pc to always encompass the Jeans length in our high density gas at the sink threshold $n = 850 \mbox{ cm}^{-3}$ (see Section~\ref{subsec:sinks}), and therefore do not employ additional Jeans refinement of gas cells in these models. 

The resulting spatial resolution is a function of density, scaling as $R_{\rm cell} \propto n^{-1/3}$, where $R_{\rm cell}$ is the radius of a mesh cell (approximating Voronoi mesh cells as spherical). We do use an additional geometric refinement condition, which affects the maximum volume of cells within a set region. Any gas cells within a galactocentric radius of 12\,kpc in the $xy$ plane and within a vertical distance above or below the disc of 1\,kpc (the $z$ plane) are constrained to have a maximum volume of $10^{6}\,\mathrm{pc^{3}}$. This is equivalent to a radius of $\sim$ 62\,pc, assuming spherical cells. This results in a minimum level of resolution within the disc, and allows us to increase the maximum volume of cells outside the region, where a minimum resolution is unnecessary. We note that our resulting resolution, while high in dense regions, is coarse in diffuse regions. This is not ideal for modelling of magnetic fields, as the field is sensitive to high velocity gradients and strong vorticity, which are strong effects in diffuse gas \citep{Gent2021Small-scaleTurbulence}. Even our minimum resolution of $\sim$ 62\,pc within the disc is, however, approaching the criterion of $\Delta x < 80$\,pc that \citet{Martin-Alvarez2022TowardsConvergence} found was required to follow dynamo action. Our focus in this work is galactic star formation, though, so this is a necessary trade-off for highly resolved dense gas. The variation in spatial resolution as a function of density for the MHD simulation is presented in Figure~\ref{fig:resolution}. In both simulations, at the sink threshold number density of $n = 850\mbox{ cm}^{-3}$, cells are just over parsec-sized, with radii of $\sim$ 2.5\,pc. 


The temporal resolution in these simulations is also variable; the time-step varies depending on the local conditions rather than being globally determined. Variable resolution allows {\sc{Arepo}} to be much more computationally efficient than a code with a constant global time-step when dealing with a problem with such a large spatial and temporal dynamic range as the ISM dynamics of a galaxy. 

\begin{figure*}
    \centering
	\includegraphics[width=1.0\textwidth]{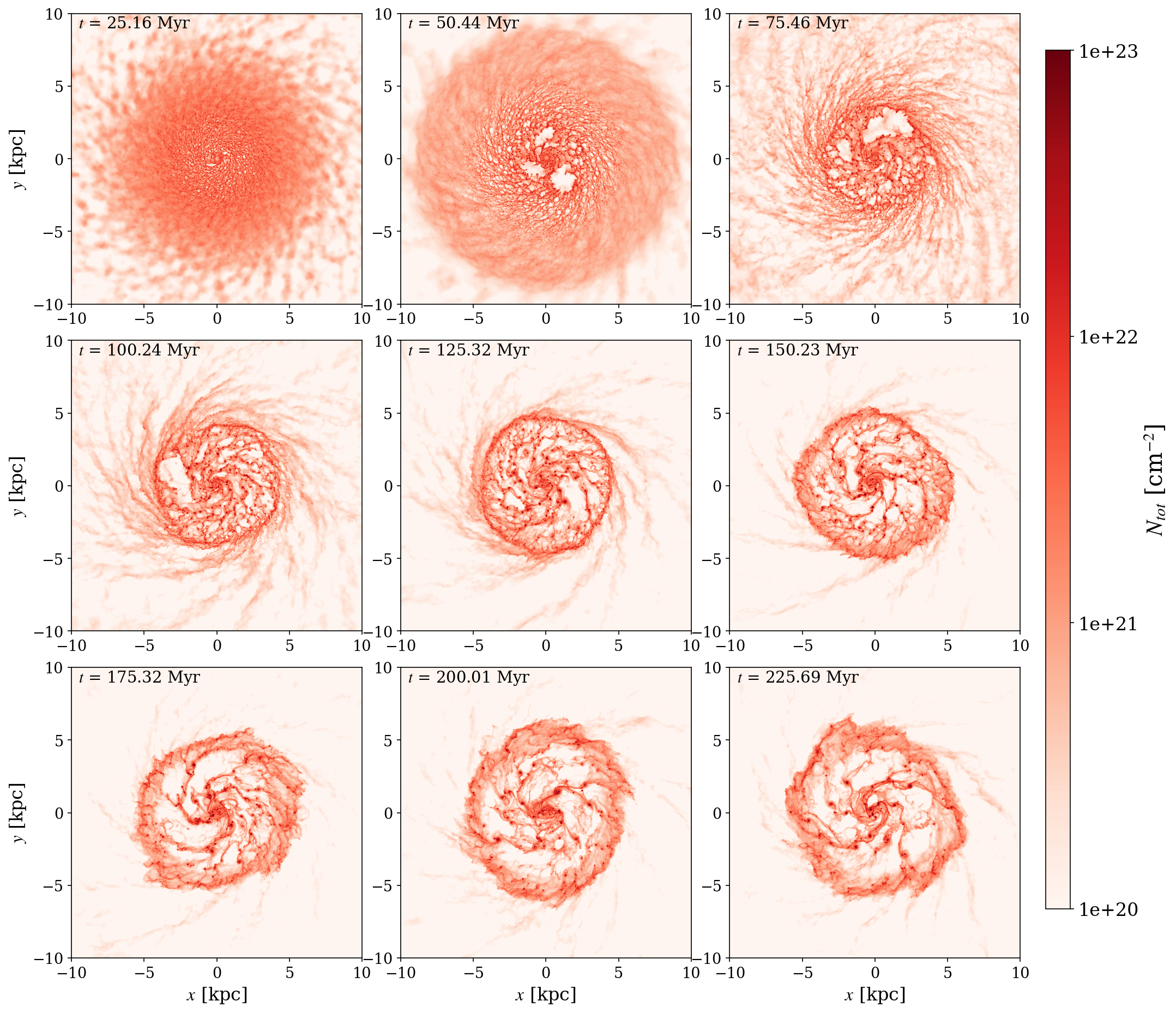}
    \caption[]{Face-on total gas column density $N_{\rm tot}$ for the MHD simulation at $\sim$ 25\,Myr intervals, illustrating the morphological evolution over 225\,Myr. These projections are column integrals along sightlines for each projected pixel, calculated by ray tracing through the Voronoi grid.}
    \label{fig:MHD_col_grid_KSP}
\end{figure*}

\begin{figure*}
    \centering
	\includegraphics[width=1.0\textwidth]{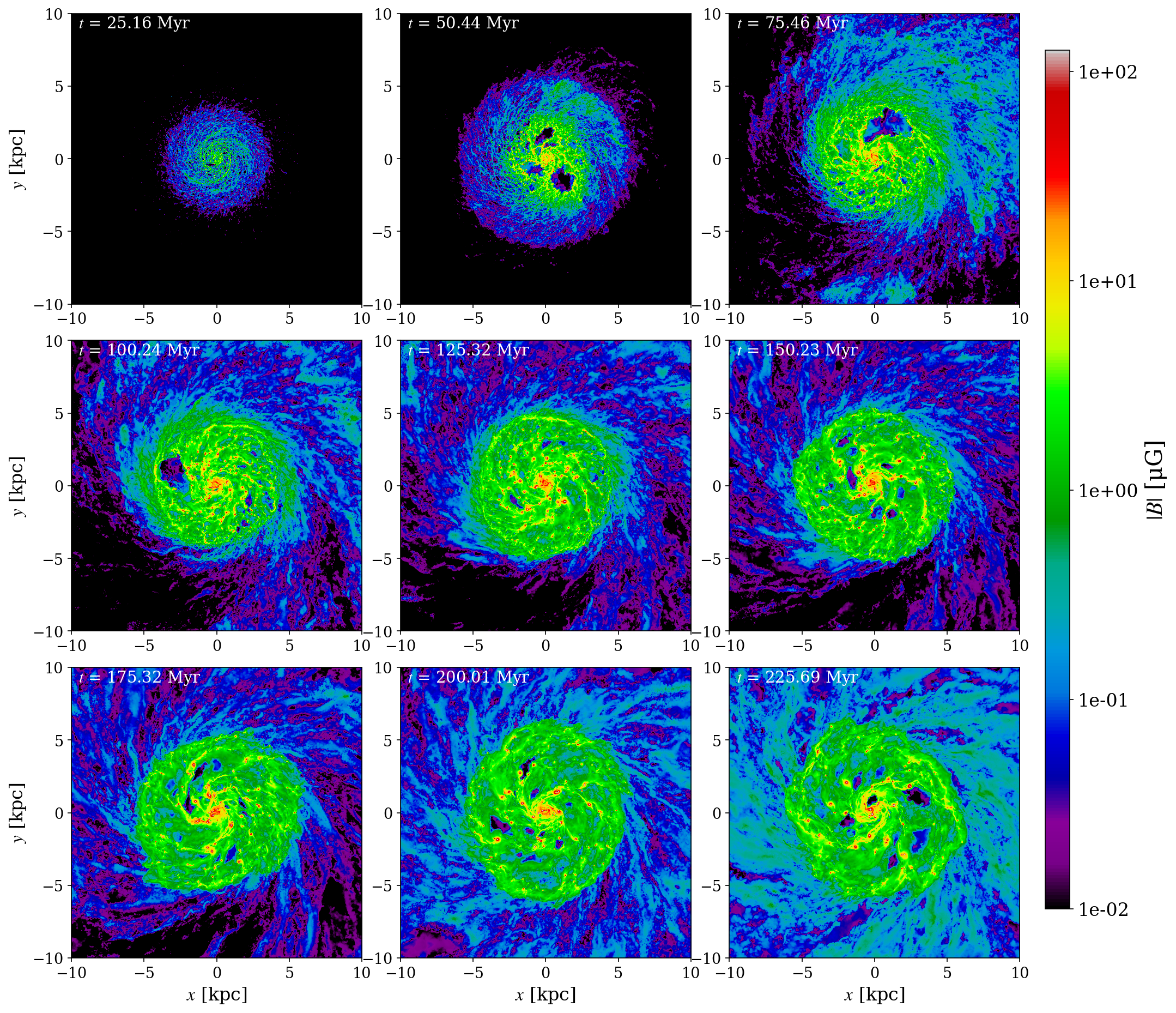}
    \caption{Face-on projections of the absolute magnetic field strength,  $|B| = (B_x^2 + B_y^2 + B_z^2)^{1/2}$, calculated as mass-weighted column integrals, shown for the same intervals as in Figure~\ref{fig:MHD_col_grid_KSP}. In this way, the gas column density and absolute magnetic field strengths across the disc can be visually compared.}
    \label{fig:mag_grid}
\end{figure*}

\begin{figure*}
    \centering
    \includegraphics[width=\textwidth]{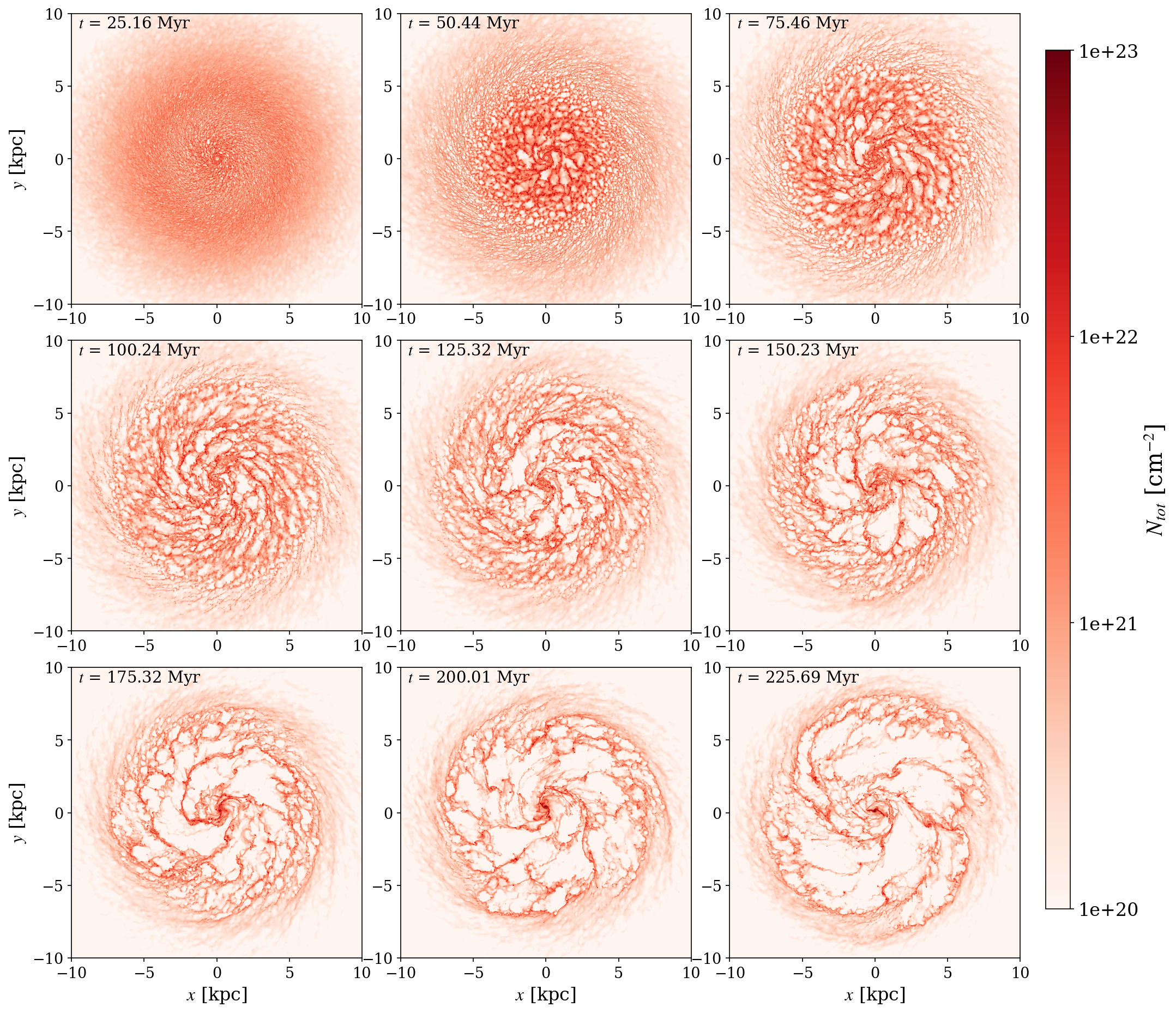}
    \caption{Face-on total gas column density $N_{\rm tot}$ for the HD simulation, for the same time intervals as  Figure~\ref{fig:MHD_col_grid_KSP}.}
    \label{fig:HD_col_grid}
\end{figure*}

\subsubsection{Sink Particles}
\label{subsec:sinks} 

While {\sc{Arepo}} allows us to run simulations over a broad range of length scales, it remains computationally unfeasible to run galaxy-scale simulations that resolve the small scales and extremely high densities necessary to fully capture the star formation process. Instead, we use collisionless sink particles as a sub-grid model, as done in many other studies of star formation (see for example \citealt{Bate1995ModellingSystems}, \citealt{Federrath2010ModelingSPH}, \citealt{Tress2020SimulationsGalaxy}, and \citealt{Whitworth2021IsMetallicities}). In our simulations, these accreting sink particles represent clusters of stars, and the chosen sink parameters reflect this. Values for the critical density $\rho_{\rm c}$, the accretion radius $r_{\rm acc}$, the gravitational softening length, sink efficiency $\epsilon_{\rm SF}$, and scatter radius $r_{\rm sc}$ are given in Table~\ref{tab:sink_params_KSP}. A fully comprehensive description of the sink particle implementation in {\sc Arepo} can be found in \cite{Tress2020SimulationsGalaxy}, but we highlight the important features here for completeness. 


We set a critical density threshold $\rho_{\rm c}$ above which sink particles may form. Sinks may only form, however, if the following conditions are satisfied:
\begin{enumerate}
    \item The gas flow in the region is converging. To establish this, it is required not only that the velocity divergence $\nabla \cdot \boldsymbol{v} < 0$ but also that the divergence of the acceleration $\nabla \cdot \boldsymbol{a} < 0$.
    \item The region is located at a local minimum of the potential.
    \item The region is neither situated within the accretion radius of another sink nor will move within the accretion radius of another sink in a time less than the local free-fall time.
    \item The region is gravitationally bound, i.e.\ $U > 2(E_{\rm k} + E_{\rm th})$, where $U = GM^2/r_{\rm acc}$ is the gravitational energy of the mass $M$ within the accretion radius, $E_{\rm k} = (1/2)\Sigma_i m_i \Delta v_i^2$ is the total kinetic energy of all gas particles within the accretion radius with respect to the centre of collapse, and $E_{\rm th} = \Sigma_i m_i e_{{\rm th},i}$ is the total internal energy of the same region. Here $e_{{\rm th},i}$ is the specific internal energy (per unit mass) for each gas particle $i$.
\end{enumerate}

In order to treat both our MHD and HD simulations identically, we do not alter these conditions to account for magnetic fields. However, a magnetically supported region will not form a sink particle due to the stipulation that velocity and acceleration divergences must be negative. 

Once a sink particle has formed, it can accrete mass from gas cells within the specified accretion radius $r_{\rm acc}$ for up to 2\,Myr after formation. We shut off accretion after 2\,Myr to approximate the effects of early feedback processes \citep{Chevance2021Pre-supernovaGalaxies}. Within these initial 2\,Myr, if a cell within this radius exceeds the density threshold $\rho_{c}$ and if the gas is gravitationally bound to the sink, mass will be removed from the cell and accreted onto the sink. The amount of mass removed 
\begin{equation}
    \Delta m = (\rho_{\rm cell} - \rho_{\rm c}) V_{\rm cell},
    \label{eq: sink_accretion_KSP}
\end{equation} 
where $\rho_{\rm cell}$ is the cell's initial density, and $V_{\rm cell}$ is its volume. If a gas cell passes all sink formation checks and is within the accretion radii of multiple sinks younger than 2\,Myr, mass is accreted onto whichever existing sink particle the gas cell is most strongly bound to. 


\begin{table}
    \centering
    \caption{Parameters of the sink particles: critical density $\rho_{\rm c}$, accretion radius $r_{\rm acc}$, gravitational softening length, sink efficiency $\epsilon_{\rm SF}$, and scatter radius $r_{\rm sc}$.}
    \label{tab:sink_params_KSP}
    \begin{tabular}{lccr}
    \hline$\rho_{\mathrm{c}} \,(\mathrm{cm}^{-3})$ & $850$ \\
    $r_{\mathrm{acc}}(\mathrm{pc})$ & 5.0 \\
    Softening length $(\mathrm{pc})$ & 5.0 \\
    {$\epsilon_{\mathrm{SF}}$} & 0.05 \\
    $r_{\mathrm{sc}}(\mathrm{pc})$ & 5.0 \\
    \hline
    \end{tabular}
\end{table}

Note that, as star formation is observed to still be relatively inefficient on size scales comparable to the accretion radius of our sink particles \citep[see e.g.][]{Evans2009THELIFETIMES}, we adopt a fixed star formation efficiency $\epsilon_{\rm SF} = 0.05$ (5\%) for the gas represented by the sinks. In this way, the stellar mass of a sink is only 5\% of its total mass, with the other 95\% assumed to be molecular gas. This gas mass is later returned to the ISM through supernova (SN) feedback (see Section~\ref{subsec:feedback}). Within the sink particle sub-grid model, an initial mass function (IMF) is sampled to give a stellar population. As is described in the following feedback section, stars with masses greater than $8\nobreak\,\mathrm{M_{\odot}}$ will explode as SNe at various times after the sink's formation depending on their mass. Once all the SN events have occurred, sinks convert into collisionless star particles, which do not accrete.  

\subsubsection{Feedback}
\label{subsec:feedback}

The prescription for feedback in galaxy-scale simulations strongly influences SFRs, the lifetimes of molecular clouds, and the structure of the ISM. Past work has illustrated that relying solely on galactic shear or using a random SN model---in which SN events occur stochastically based on an expected rate, in random locations---are insufficient to disrupt giant molecular clouds \citep{Walch2015TheISM}. Following \cite{Tress2020SimulationsGalaxy}, we tie feedback directly to our sink particles. With 5\% of the sink particle mass categorised as stars, a discrete stellar population is attributed to the sink by sampling from the IMF following the method described by \cite{Sormani2017AStars}. If any given sink particle contains stars more massive than $8\nobreak\,\mathrm{M_{\odot}}$ it will trigger SN events at the ends of their lifetimes. This lifetime is determined by the individual mass of each star, following Table 25.6 of \cite{Maeder2009PhysicsStars}. As the sink particles represent clusters of stars, SNe do not occur at the exact location of the sink, but randomly within the sink accretion radius. 

How energy is injected by SNe within the simulation depends on whether or not the Sedov-Taylor phase of the SN remnant is resolved. For every SN, we first compute the radius of the remnant at the end of the Sedov-Taylor phase
\begin{equation}
R_{\mathrm{ST}}=19.1\left(\frac{\bar{n}}{1 \mathrm{~cm}^{-3}}\right)^{-7 / 17} \mathrm{pc}
    \label{eq: R_ST_KSP},
\end{equation}
where $\bar{n}$ is the local mean density \citep{Blondin1998TransitionRemnants}, and we assume an SN energy of $10^{51}$\,erg and solar metallicity. The radius of injection $R_{\rm inj}$ is the radius of the smallest sphere that contains 40 cells centred on the location of the explosion. If $R_{\rm ST} > R_{\rm inj}$ so that the event is resolved, then $10^{51}$\,erg is injected thermally. If however $R_{\rm ST} < R_{\rm inj}$ so the event is unresolved, no energy is injected thermally, but rather the corresponding amount of momentum is injected. Gas within the region is fully ionised in either case. See \cite{Tress2020SimulationsGalaxy} for further details. For SN feedback in general, we set a maximum injection radius of 250\,pc. Any SN event with a minimum 40-cell sphere larger than this is skipped as a safeguard to prevent unphysical, under-resolved energy injection. In the models presented in this work, this only occurs for 1.28\% of SNe in the MHD model, and 0.85\% of SNe in the HD model. 

Mass is also returned to the surrounding ISM with each SN event. The total amount of mass returned with each event $M_{\rm ej}$ is determined by the gas mass of the sink and the number of remaining SN events, $n_{\rm SN}$, as
\begin{equation}
    M_{\rm ej} = M_{\rm sink,gas} / n_{\rm SN},
    \label{eq: mass_return}
\end{equation}
where $M_{\rm sink,gas}$ is 95\% of the total sink mass at the time when the SN event occurs. This is distributed uniformly within the injection region. 

Following this prescription, at the end of its lifetime, a sink will have returned all of the gas not used to form stars back to the ISM. This is consistent with the observation that star clusters older than 30\,Myr have no associated gas or dust \citep{Bastian2014ConstrainingGlobular}. Once the final SN for a particular sink has occurred, it is converted into a collisionless star particle. These star particles become part of the background stellar population of old stars. They do not accrete, and represent only stellar mass. If a sink particle never harbours a star greater than 8\,M$_{\odot}$ and therefore causes no SN events, it is converted to a star particle after 10\,Myr. The 95\% of its mass that does not contribute to the stellar mass is returned uniformly to the surrounding ISM, distributed among all cells within a 100\,pc radius. 

\begin{figure}
    \centering
	\includegraphics[width=0.5\textwidth]{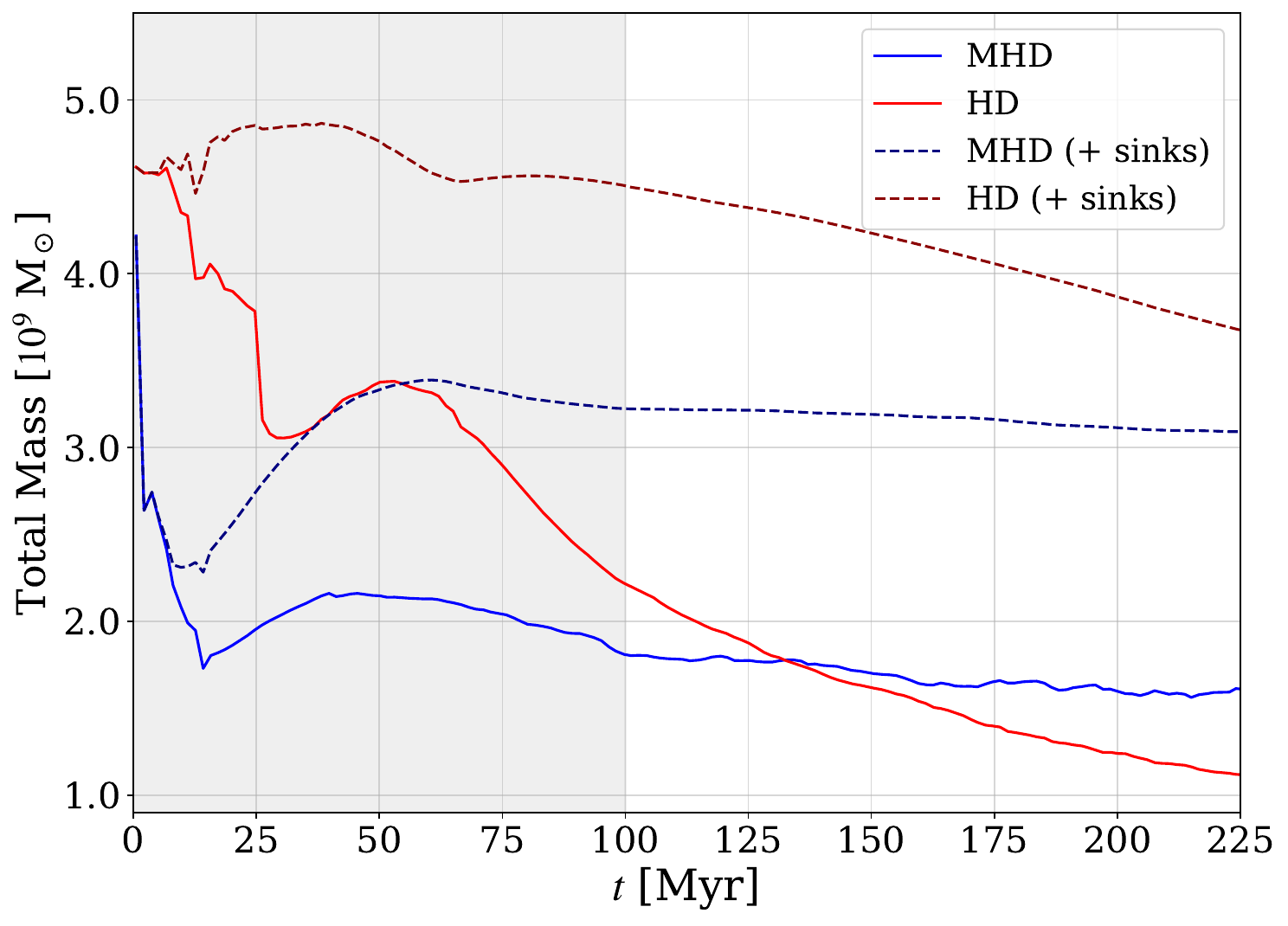}
    \caption[Gas mass of both MHD and HD discs over time]{Total gas mass of the disc of the MHD {\em (solid blue)} and HD {\em (solid red)} models over time. The {\em grey} region denotes the initialisation phase, during which the galaxies are still responding to their initial condition. We see a gradual depletion in mass, mostly resulting from star formation. We additionally plot the combined mass of gas cells and the gas in sink particles, for both the MHD {\em (dotted blue line)} and HD {\em (dotted red line)} models. As gas is converted into stars, and there is little to no gas in-flow to the disc of our isolated galaxies, the gas reservoir is eventually depleted.}
    \label{fig:gas_depletion}
\end{figure}

\subsubsection{Chemistry}
\label{subsec:chem} 
To follow the chemical evolution of the gas, we use a time-dependent, non-equilibrium chemical network, specifically the NL97 network of \citet{Glover2012}. This combines the network for hydrogen chemistry presented in \cite{Glover2007SimulatingConditions,Glover2007SimulatingConditionsb} with a simplified treatment of CO formation and destruction introduced by \cite{Nelson1997TheFields}, as was first implemented in {\sc{Arepo}} by \cite{Smith2014CO-darkGalaxies}. We assume a cosmic ray ionisation rate of $3 \times 10^{-17}{\rm s}^{-1}$ for atomic hydrogen, with the rates for other species scaled appropriately. We adopt solar metallicity, with gas phase abundances of C and O (relative to H, by number) of $1.4 \times 10^{-4}$ and  $3.2 \times 10^{-4}$, respectively. We assume a constant background far-ultraviolet interstellar radiation field at the solar neighbourhood value \citep{Draine1978PhotoelectricGas} of $G_{0} = 1.7$ \citep{Habing1968TheA}. CO and H$_2$ self-shielding and the shielding from dust absorption are modelled using the {\sc{TreeCol}} algorithm developed by \cite{Clark2012TreeCol:Simulations}. When computing the shielding of a particular Arepo cell, we assume that only gas within a distance of 30\,pc contributes, as in the solar neighbourhood, the typical distance to the nearest O or B star is approximately 30\,pc \citep{Maiz-Apellaniz2001TheHipparcos, Reed2003CatalogStars, Smith2014CO-darkGalaxies}. 

The radiative and chemical heating and cooling of the gas is also computed simultaneously with the chemical evolution using a detailed atomic and molecular cooling function as detailed by \citet{Clark2019TracingEmission}. The treatment of cooling at high temperatures ($T \gg 10^{4}$K) is important for accurate modelling of SN feedback. Cooling at these temperatures from atomic hydrogen is calculated using the non-equilibrium H and e$^{-}$ abundances from the chemistry, while cooling from He and other metals assumes collisional ionisation equilibrium and uses values given by \cite{Gnat2012ION-BY-IONEFFICIENCIES}. Strong adiabatic cooling at particular resolutions can artificially produce extremely low temperatures. To account for this, we introduce a temperature floor of 20\,K, which also prevents extremely low Jeans masses.

\subsection{Initial Conditions}
\label{subsec:init} 

\begin{table}
	\centering
	\caption{Parameters of the isolated galaxy initial condition components from \citep{Tress2020SimulationsGalaxy}}
	\label{tab:IC_params}
	\begin{tabular}{lccr} 
		\hline
		   & $M$ (M$_{\odot}$) & $a \mbox{ or }h_R$ (kpc) & $h_{z}$ (kpc)\\
		\hline
		Dark matter halo & $6.04 \times 10^{11}$ & 28.7 & -\\
		Bulge & $5.30 \times 10^{9}$ & $9.03 \times 10^{-2}$ & -\\
		Stellar disc & $4.77 \times 10^{10}$ & 2.26 & 0.3\\
		Gas disc &  $5.30 \times 10^{9}$ & 2.26 & 0.3\\
		\hline
	\end{tabular}
\end{table}

The simulations presented in this work start from the same initial conditions as those of the isolated model presented by \cite{Tress2020SimulationsGalaxy}. Here we recap the most important details.

\begin{figure*}
    \centering
	\includegraphics[width=0.95\textwidth]{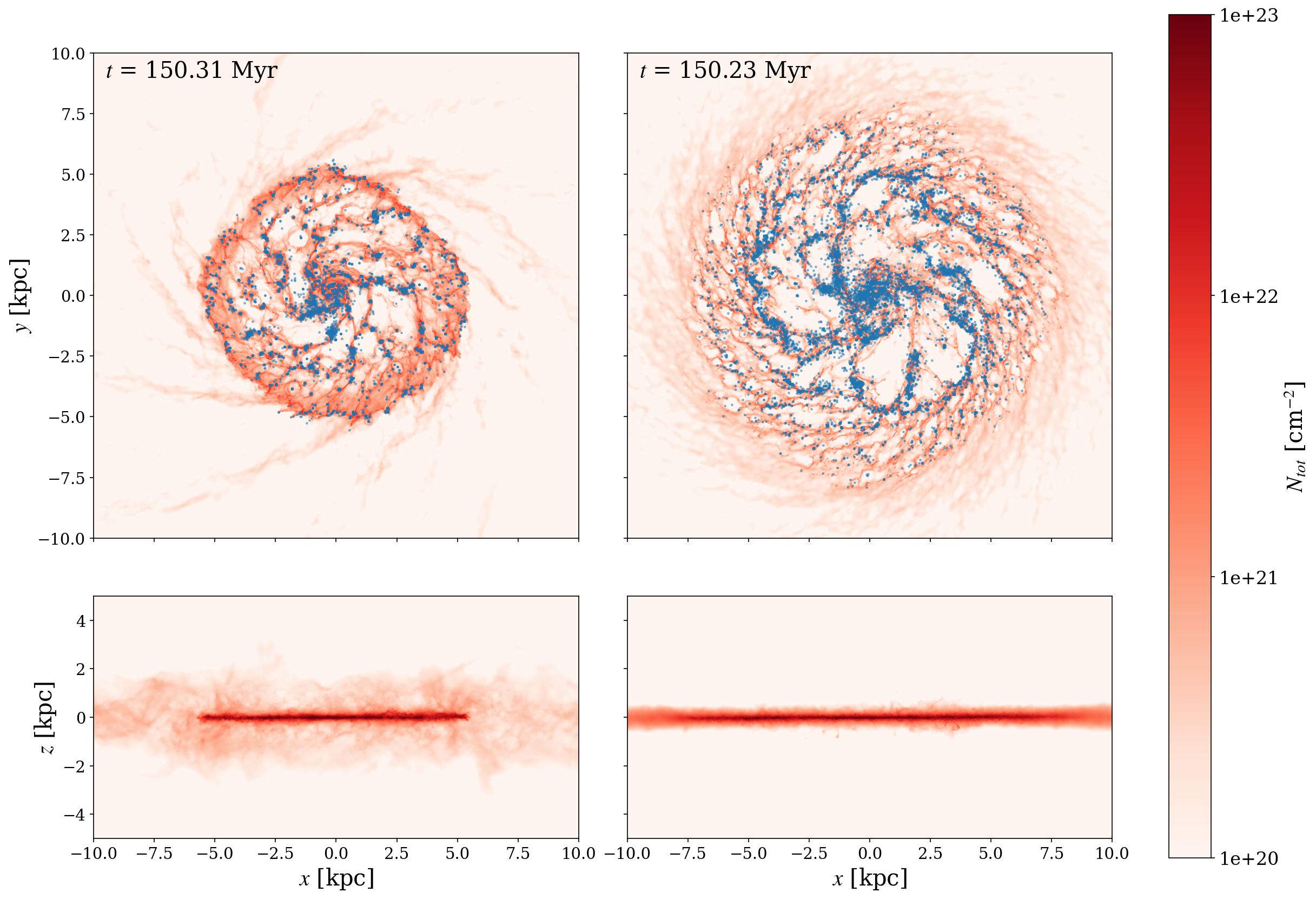}
    \caption[]{Face-on ({\em top)} and edge-on {\em (bottom)} total column densities of the MHD {\em (left)} and HD {\em (right)} simulations at $\sim$ 150\,Myr. In the face-on panels, sink particle positions are overplotted in blue with partial transparency, so darker blue points indicate the presence of multiple sink particles in a small area. At this time, the MHD simulation has a total of 17652 sink particles, with a corresponding stellar mass in sinks of $\sim$ 78.7 $\times 10^{6}\,\mathrm{M_{\odot}}$, while the HD has 35482, and a corresponding stellar mass in sinks of $\sim$ 137.6 $\times 10^{6}\,\mathrm{M_{\odot}}$. In the  edge-on panels, the sink particles have been omitted for clarity.}
    \label{fig:proj_with_sinks}
\end{figure*}

The galaxy initial condition includes a dark matter halo, a stellar bulge, a stellar disc, and a gaseous disc. Both the stellar bulge and dark matter halo follow a spheroidal \cite{Hernquist1990AnBulges} profile 
\begin{equation}
    \rho_{\rm sp}(r) = \frac{M_{\rm sp}}{2\pi} \frac{a}{r(r+a)^{3}},
    \label{eq: hernquist_KSP}
\end{equation}
where $r$ is the spherical radius, $a$ is the scale-length of the relevant spheroid
, and $M_{\rm sp}$ is its mass.
The stellar and gas disc follow double exponential density profiles given by
\begin{equation}
    \rho_{\rm disc}(R,z) = \frac{M_{\rm disc}}{4 \pi h_{z} h^{2}_{R}} \mbox{sech}^{2} \frac{z}{2h_{z}} \exp \left(- \frac{R}{h_{R}}\right)\;,
    \label{eq: dbl_exp}
\end{equation}
where $R$ and $z$ are the cylindrical radius and height and $h_{z}$ and $h_{R}$ are the scale-height and scale-length of the disc, respectively. The chosen parameters to construct these profiles are given in Table~\ref{tab:IC_params}; for  justifications for these values see \cite{Tress2020SimulationsGalaxy}. As in that study, we assume fully atomic gas at initialisation, set at T = 10$^4$ K, and assume the ISM to be solar metallicity throughout.

For the MHD case, the field is initialised at a starting value of $B_{\rm init} = 0.02\,\upmu\mathrm{G}$. This value is low enough to allow the turbulent dynamo to amplify the field self-consistently \citep[e.g.][]{Korpi-Lagg2024ComputationalGalaxies}, and high enough to reach a stable configuration in a reasonable time frame. 
\citet{Whitworth2022MagneticGalaxies} further discuss the effects of initial seed field strength and orientation. 
The initial direction of the field is toroidal with $x$ and $y$ components 
\begin{align}
B_{x} &= -B_{\rm init} \sin(\theta), \label{eq: tor_field_x} \\
B_{y} &= B_{\rm init} \cos(\theta) \label{eq: tor_field_y}
\end{align}
at all altitudes.
The $z$ component of the magnetic field is zero at initialisation. Note that we quote volume-weighted measures of the magnetic field strength in our results.

\section{Results}
\label{sec:results} 

We start by visualising the overall evolution of both models, including the growth and evolution of the magnetic field, before studying the difference between the star formation and gas phase with and without fields. Snapshots of the simulations are presented, and videos of their full evolution are also available\footnote{Full evolution videos available as mp4 files attached to this publication, and also at \url{https://kammybogue.github.io/bogue_isolated_galaxy_sims.html}}. Unless otherwise stated, our analysis of the gas in the models pertain only to the gas particles exclusively, i.e neglecting the gas locked in sink particles.

\subsection{Morphological Evolution}
\label{subsec:morph_evo}

We run both simulations to $\sim$ 225\,Myr. As will be discussed later, beyond this point the HD model has lost most of its gas, so we do not run further. In Figure~\ref{fig:MHD_col_grid_KSP} we show the morphological evolution of the MHD model, through the face-on total gas column density ($N_{\rm tot}$). At intermediate times, shortly after the initialisation phase, we see a ring-like structure forming toward the edge of the disc that is not seen in the HD equivalent. This becomes less sharply defined at late times. This is a clear difference between the MHD and HD simulations. While ring-shaped patterns are relatively common in local galaxies \citep[$\sim$ 20\% of them,][]{Fernandez2021PropertiesStructures}, we cannot be sure that this is a physical result. In test MHD models, with varying sink particle size, ring-like overdensities did form consistently but were more quickly disrupted with smaller sink particles. The dispersal of this structure in these tests could be a combination of more sources of feedback or differences in the field strength in dense gas. However, the ring structure could also be due to the disc being initially out of hydrostatic equilibrium because of the additional pressure of the magnetic field. 

In dense regions of the MHD simulation, large extended clumps of gas are seen, which coincide with regions of high star formation. Similar features are seen in a number of other MHD models \citep[see for example][]{Kortgen2018TheGalaxies}. However, we are once again cautious about these features, as we note that they may be affected by resolution of sink particles. 


In Figure~\ref{fig:mag_grid} we show the evolution of the galactic magnetic field at the same times as in Figure~\ref{fig:MHD_col_grid_KSP}. We see the action of the turbulent dynamo in the growth of the magnetic field magnitude from its initial seed value. Amplification is more rapid in the inner galaxy at first. The magnetic field becomes more and more extended, until the majority of the disc has field strengths on the order of a few microgauss. The field structure aligns closely with the column densities of Figure~\ref{fig:MHD_col_grid_KSP} as expected given our ideal MHD assumption. The large extended clumps of gas seen in $N_{\rm tot}$ correspond to regions with an above average field strength, particularly in their centers. 

The equivalent morphological evolution for the HD model is presented in Figure~\ref{fig:HD_col_grid}. There are clear differences to the MHD evolution. The spiral structure is more flocculent, the peak $N_{\rm tot}$ values are lower. Voids are seen in both models, but there is no evidence of the ring-like structure seen in the MHD model. At early times, the bubbles caused by SN feedback are more clearly evident than in the MHD model. At late times, significantly larger voids appear, likely because of gas depletion. Visually, the HD disc also appears to extend further than the MHD disc, which is more compact. We quantify this for our fiducial 150\,Myr snapshot by calculating the radius within which 75\% of the gas mass is contained. For the MHD model, we find this radius to be 5.1\,kpc, as compared to 7.4\,kpc for the HD model.

We calculate the total mass of gas in the disc by summing all gas cell masses within a set cylindrical region of radius 10\,kpc and extending $z = \pm$0.25\,kpc from the midplane of the disc. In Figure~\ref{fig:gas_depletion} we show the variation in this total disc gas mass over time. For clarity, we also illustrate the total gas mass when including the dense gas within sink particles, by adding 95\% of sink particle masses to the total gas mass. During the first 100\,Myr, the galaxies are still responding to their initial condition. We refer to this as the {\em initialisation phase} hereafter. There is an initial rapid fall in gas mass due to the first burst of sink particle formation, which is exacerbated in the MHD case by gas flowing vertically out of our measurement region into the halo. This is discussed in more detail in the following sections.

Figure~\ref{fig:gas_depletion} shows that the total gas mass in both discs is comparable between 125--150\,Myr, so for analyses integrated over time we use this period. For other analyses we focus on the 150\,Myr data. The continual drop in gas mass is not unexpected: when gas becomes gravitationally unstable, it is consumed by sinks, which continue to accrete gas from their surroundings. At the end of a sink's lifetime, 95\% of the mass is returned to the ISM, but 5\% is permanently locked up into collisionless star particles. Given that these are isolated galaxy models, there is no gas replenishment besides the galaxy's own atmosphere, and so the total disc gas mass decreases over time, particularly in the HD case which does not form a substantial atmosphere (see Figure\ \ref{fig:proj_with_sinks}, bottom panels). Hence, we do not continue running to later times, as the models cease to be representative of nearby spiral galaxies when they are so gas depleted. For this study, we rather focus on treating both models in the same manner, and analysing the same time period, in order to compare the two simulations robustly.


We compute the average depletion time for each model between 125--150\,Myr as the average mass in the gas reservoir over this period divided by the average star formation rate. This includes gas in sink particles, so that
\begin{equation} \tau_{\rm dep} \;=\; \frac{\displaystyle \left\langle M_{\rm gas} \;+\; 0.95\,M_{\rm sink}\right\rangle} {\displaystyle \left\langle \mathrm{SFR}\right\rangle}, \label{eq:tdep} \end{equation}
where $M_{gas}$ is the mass of gas cells within the disc, $M_{sink}$ is the mass of sink particles, and $\mathrm{SFR}$ is the star formation rate. This gives an average value for $\tau_{\rm dep}$ of 655\,Myr for the MHD case and 496\,Myr for the HD. Our galaxy depletion times are shorter than the average star-forming main-sequence galaxy, likely due to ineffective feedback, in particular a lack of early stellar feedback \citep{Andersson2024Pre-supernovaefficiency}.

\begin{figure}
    \centering
    \includegraphics[width=0.5\textwidth]{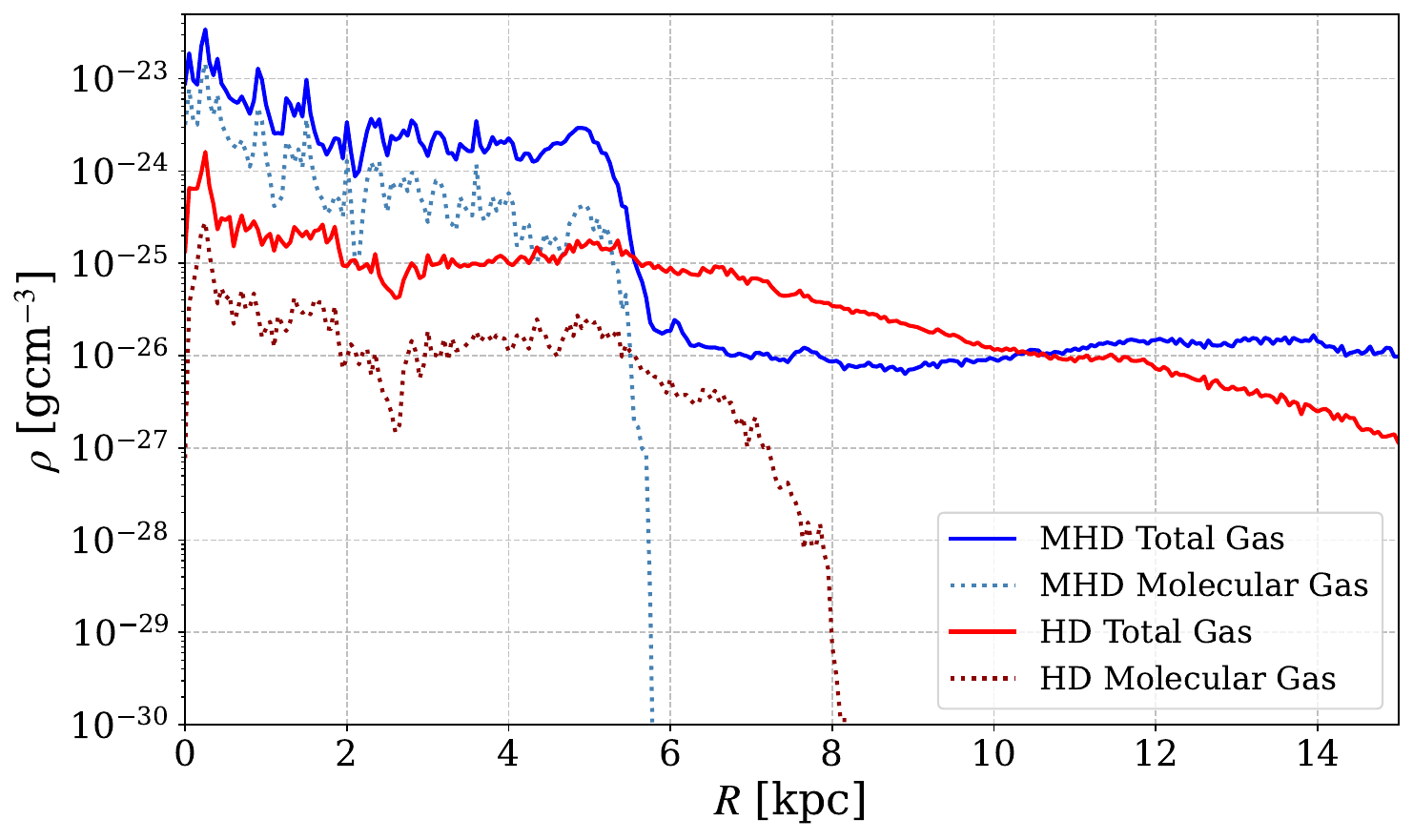}
    \caption{Volume-weighted radial density profiles for total gas density {\em (solid line)} and the molecular gas density {\em (dashed line)}, for the MHD {\em (blue)} and HD {\em (red)} models at 150\,Myr. The difference in extent of unbound molecular gas between the MHD model and the HD model is clear: the MHD molecular disc is smaller than the HD molecular disc.}
    \label{fig:radial_prof}
\end{figure}

\begin{figure}
    \centering
    \includegraphics[width=0.5\textwidth]{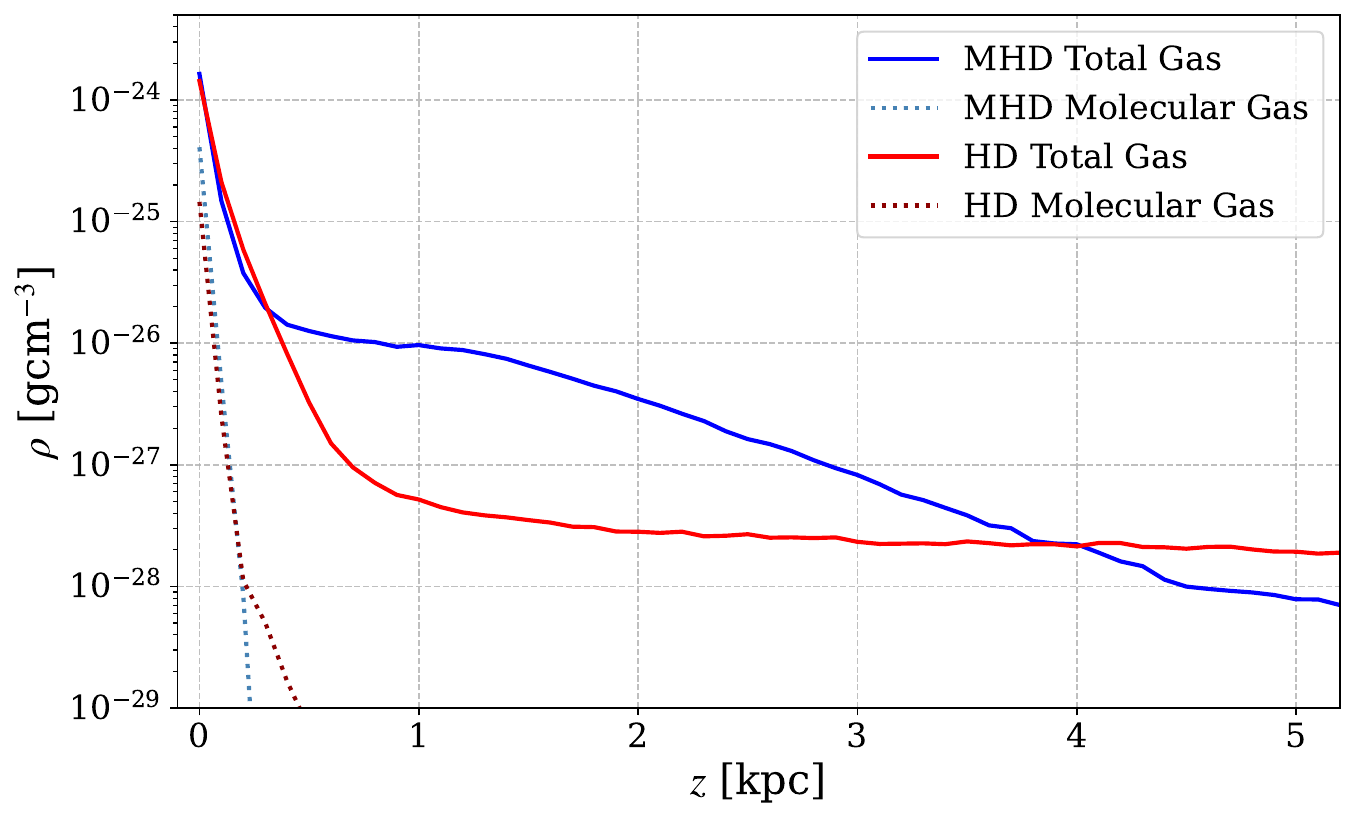}
    \caption[]{Volume-weighted vertical gas density profiles in both total {\em (solid line)} and molecular gas {\em (dashed line)} for the MHD {\em (blue)} and HD {\em (red)} models at 150\,Myr.}
    \label{fig:z_prof}
\end{figure}

In Figure~\ref{fig:proj_with_sinks} we show face-on and edge-on views of the total gas column density $N_{\rm tot}$ at $\sim$ 150\,Myr. In the face-on panels (top), we also represent the sink particle distribution. Comparing the two models visually, star formation is clumpier and less radially extended in the MHD case than in the HD case. In the edge-on view (bottom panels), we see a diffuse atmosphere above and below the disc forms in the MHD case, but not in the HD case. Similar behaviour was seen in the dwarf galaxy models of \citet{Whitworth2022MagneticGalaxies} and the stratified boxes of \citet{Girichidis2018TheClouds}. The difference in radial extent can also be seen clearly from the edge-on perspective, with a distinct decrease in $N_{\rm tot}$ in the MHD case at a radius of $\sim$ 5\,kpc. The HD model displays a much smoother fall-off of $N_{\rm tot}$. 

To further investigate the variation in radial gas distribution, we calculate the average volume-weighted gas density in radial bins. We use radial bins of width of 0.05\,kpc, from 0 to 15\,kpc. The $z$ height of the sample volume is kept constant, at $\pm$0.25\,kpc above and below the plane. The radial profile is plotted in Figure~\ref{fig:radial_prof}, which shows the density profiles for both the total gas and molecular gas. The lack of molecular gas at radii exceeding 5--8 kpc means that atomic gas is completely dominant at large radii. The total gas density follows a roughly exponential profile, with a dip in the MHD case at $\sim$ 5\,kpc and a smoother reduction in the HD case. In the molecular gas, there are much sharper drops in density.
 
To quantify the disc thickness, we calculate the volume-weighted vertical gas profile, which is shown in Figure~\ref{fig:z_prof}. We use vertical bins of width 0.1\,kpc inside a fixed radius of 10\,kpc. In both simulations, the molecular gas density decreases rapidly, and there is likewise a large decrease in the atomic gas density within 0.5\,kpc of the midplane. Above a height of $\sim 0.5$\,kpc the models differ. In the MHD case there exists an excess of diffuse atomic gas just as seen in Figure~\ref{fig:proj_with_sinks}. This atmosphere is likely a result of the rapidly dynamo-generated poloidal magnetic fields in the MHD model. As the gas is tied to the field, the easiest direction for diffuse gas to flow is along the poloidal field lines, above and below the disc. 

\subsection{Magnetic Field Evolution}
\label{subsec:bfield3} 

\begin{figure}
    \centering
 	\includegraphics[width=0.5\textwidth]{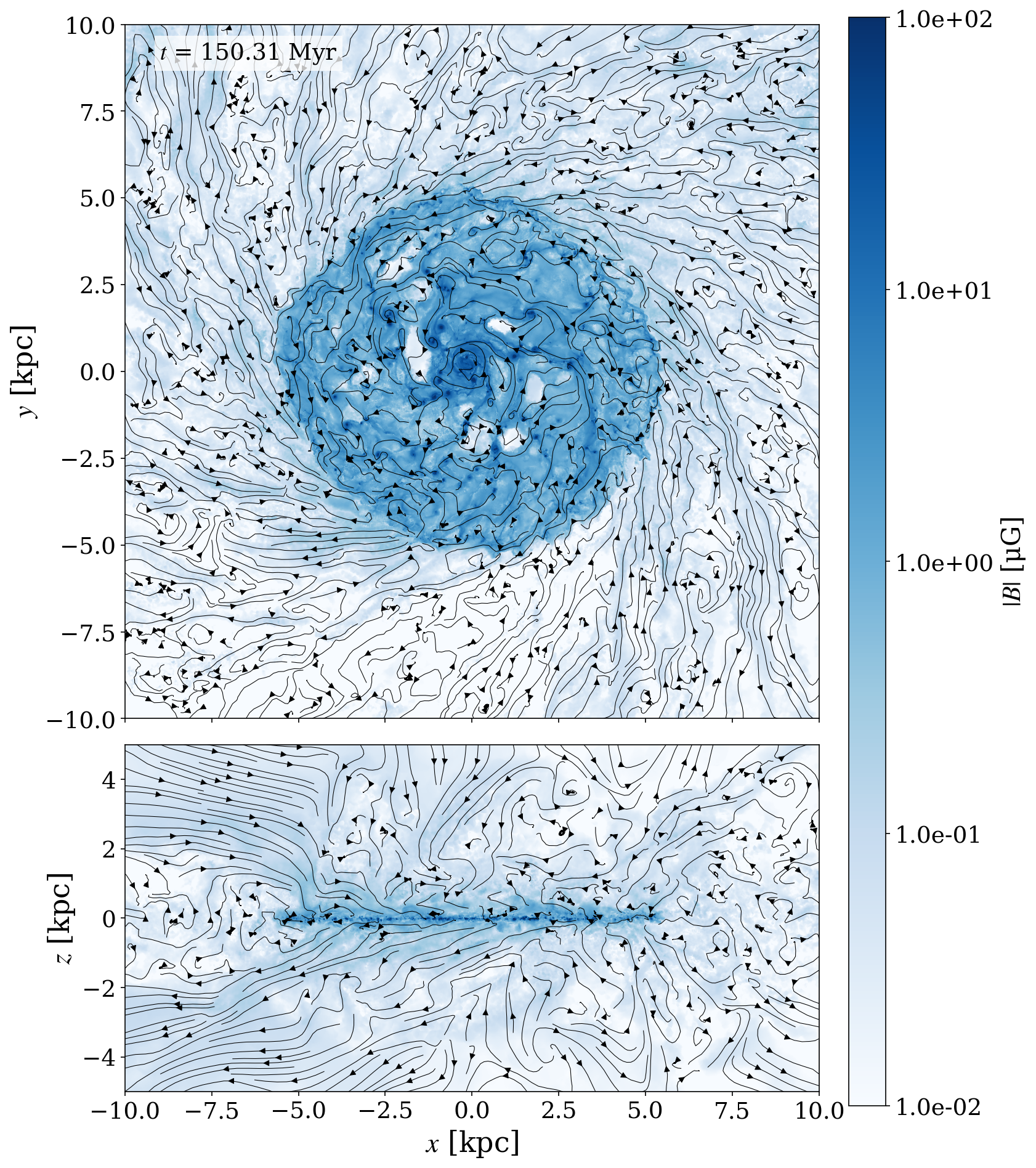}
     \caption[Magnetic field projection with streamlines]{The magnetic field magnitude $|B|$ of the MHD model in mass-weighted projection at $\sim$ 150\,Myr, as in Figure~\ref{fig:mag_grid},  in face-on {\em (top)} and edge-on {\em (bottom)} view. Over-plotted streamlines (black arrows) indicate the directionality of the field perpendicular to the line of sight in each view. At initialisation, the fully toroidal field would appear in these projected streamlines as concentric circles in the face-on view, and as horizontal lines in the edge-on view.}
     \label{fig:streamlines}
\end{figure}

\begin{figure}
    \centering
 	\includegraphics[width=0.45\textwidth]{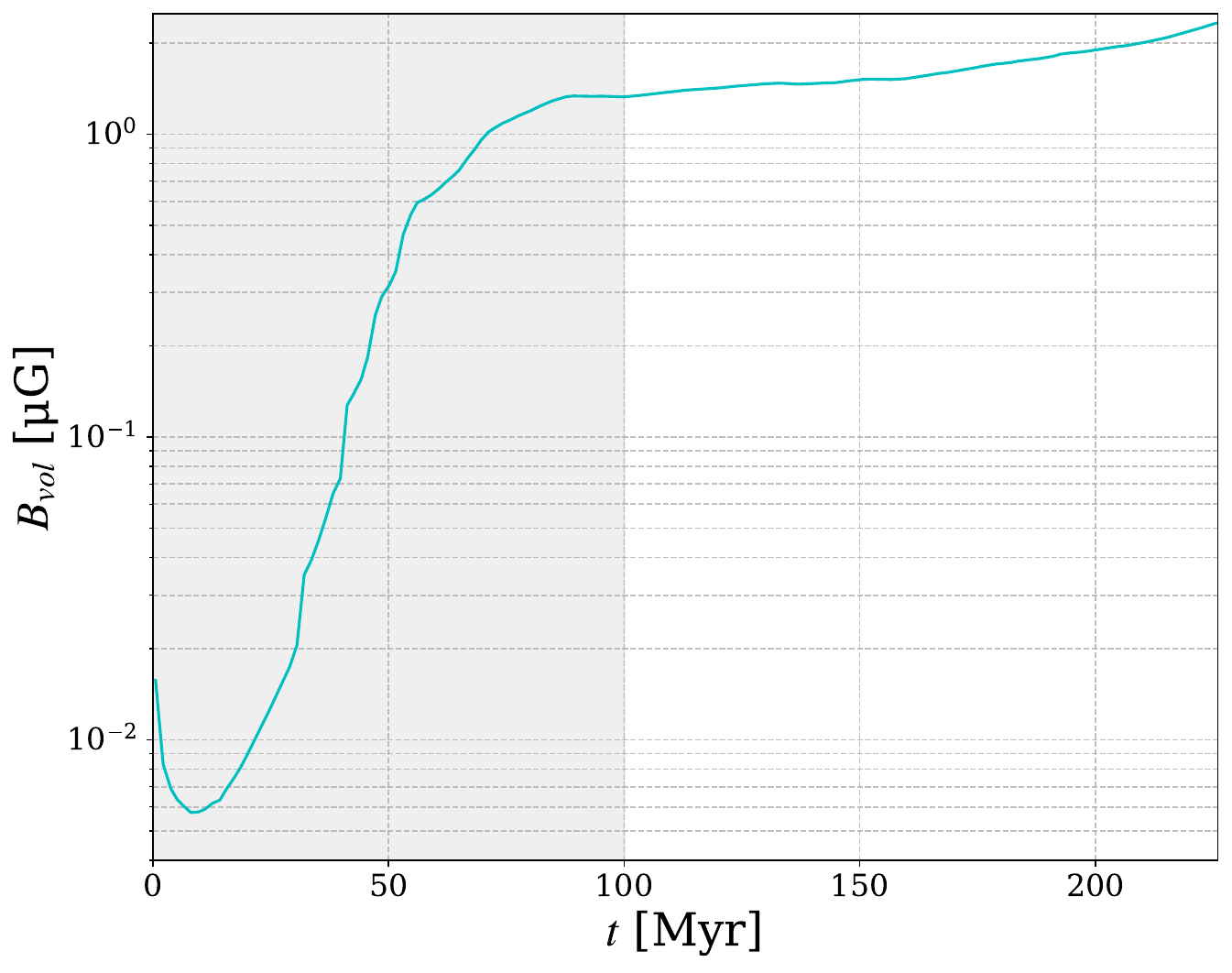}
     \caption[Amplification of the magnetic field over time]{Volume-weighted absolute magnetic field strength of the disc, showing the amplification of the field. Our field amplifies from an initial strength of $0.02\,\upmu\mathrm{G}$ to $\sim2.5\,\upmu\mathrm{G}$.}
     \label{fig:vol_weight_b}
\end{figure}

\begin{figure}
    \centering
    \includegraphics[width=0.5\textwidth]{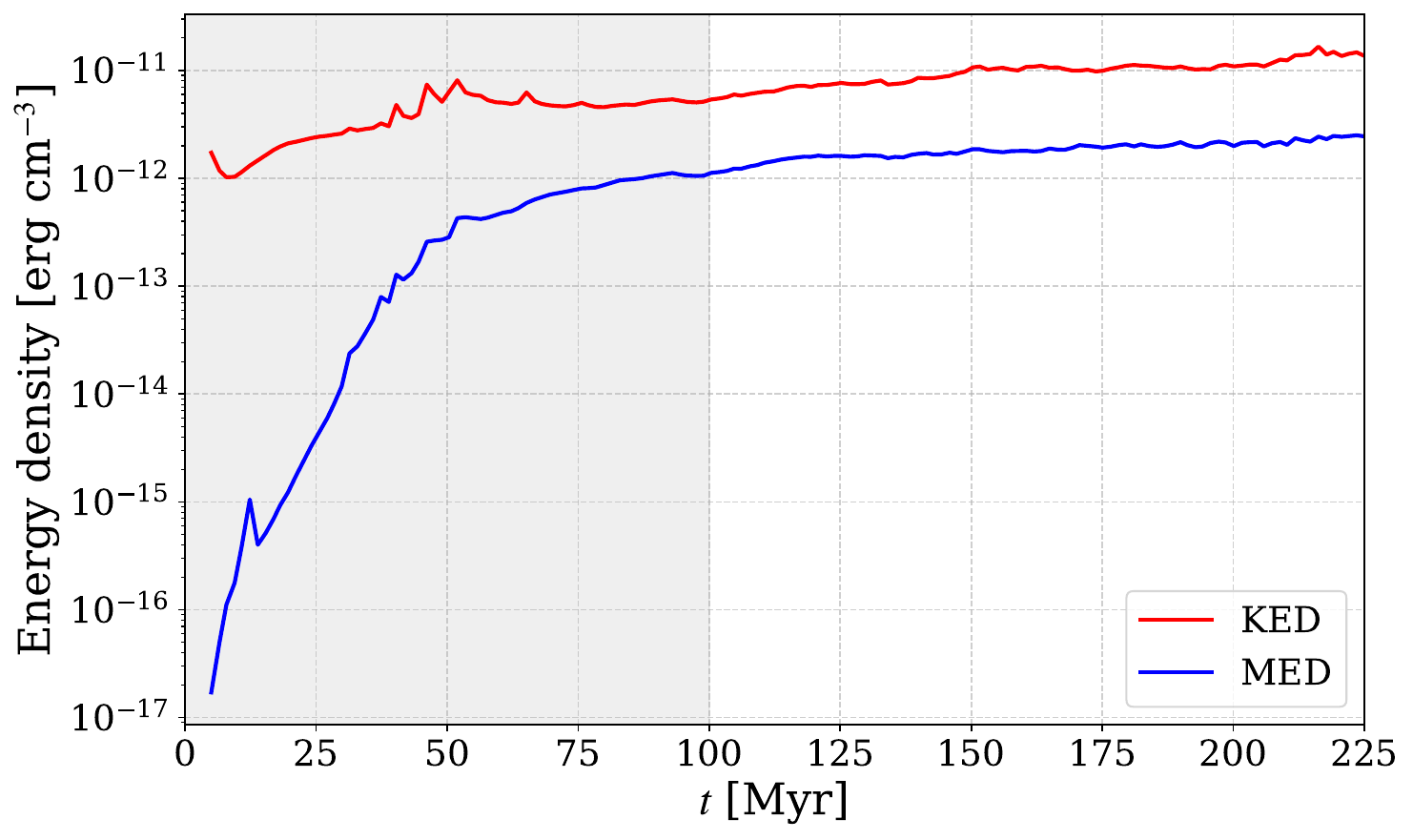}
    \caption[]{The magnetic {\em (blue)} and kinetic {\em (red)} energy density of the MHD model. Exponential growth in energy density is characteristic of dynamo activity.}
    \label{fig:MED_KED_KSP_time}
\end{figure}

The MHD model we present is initialised with a small purely toroidal seed field, with an absolute volume-weighted magnetic field strength of 0.02\,$\upmu$G. In Figure~\ref{fig:streamlines} we show the strength and directionality of the magnetic field of the MHD model at 150\,Myr, in both a face-on (top) and edge-on (bottom) view. The field strength is calculated as the mass-weighted column integral as in Figure~\ref{fig:mag_grid}. At initialisation, the field lines appear as concentric circles in the face-on view and straight horizontal lines in the edge-on view, but there is a rapid departure from this perfectly symmetric morphology. This departure is caused by SN driven turbulence and gravitational instability of the gas disc. The field correlates with the gas, and traces the outside of bubbles. We defer any further investigation into the orientation of the magnetic field and its correlation to gas structures to future work. There is also evidence of winding of the field around dense regions of gas, which is further discussed in Section~\ref{sec:discussion}. 

In Figures~\ref{fig:vol_weight_b} and~\ref{fig:MED_KED_KSP_time}, we show the average volume-weighted magnetic field strength and the magnetic energy density as a function of time \citep[computed following][]{Whitworth2022MagneticGalaxies}, integrated over the disc inside a radius of 10\,kpc and height of $\pm 0.25$\,kpc. The magnetic and kinetic energy densities are first calculated in radial bins, with bin widths of 0.05\,kpc and a constant total vertical height of $\pm 0.25$\,kpc, as in the radial profile of density. The magnetic energy density for each radial annulus is calculated as the volume-weighted average of the magnetic energy density of every cell within that annulus, given as $|B|^2 / 8\pi$, where $|B|$ is the absolute magnetic field. The kinetic energy density is computed following \citet{Beck2015MagneticGalaxies}, by integrating $(1/2)\rho \sigma^2$, where $\sigma$ is the mass-weighted velocity dispersion of the cylindrical velocity components of the gas, $v_{r}$, $v_{\theta}$ and $v_{z}$. While $v_{\theta}$ will include some contribution from shear, we find this contribution on average to be much smaller than from turbulence, and that $v_{\theta}$ has a much smaller contribution to the overall $\sigma$ than $v_{r}$ and $v_{z}$.




\begin{figure}
    \centering
	\includegraphics[width=0.5\textwidth]{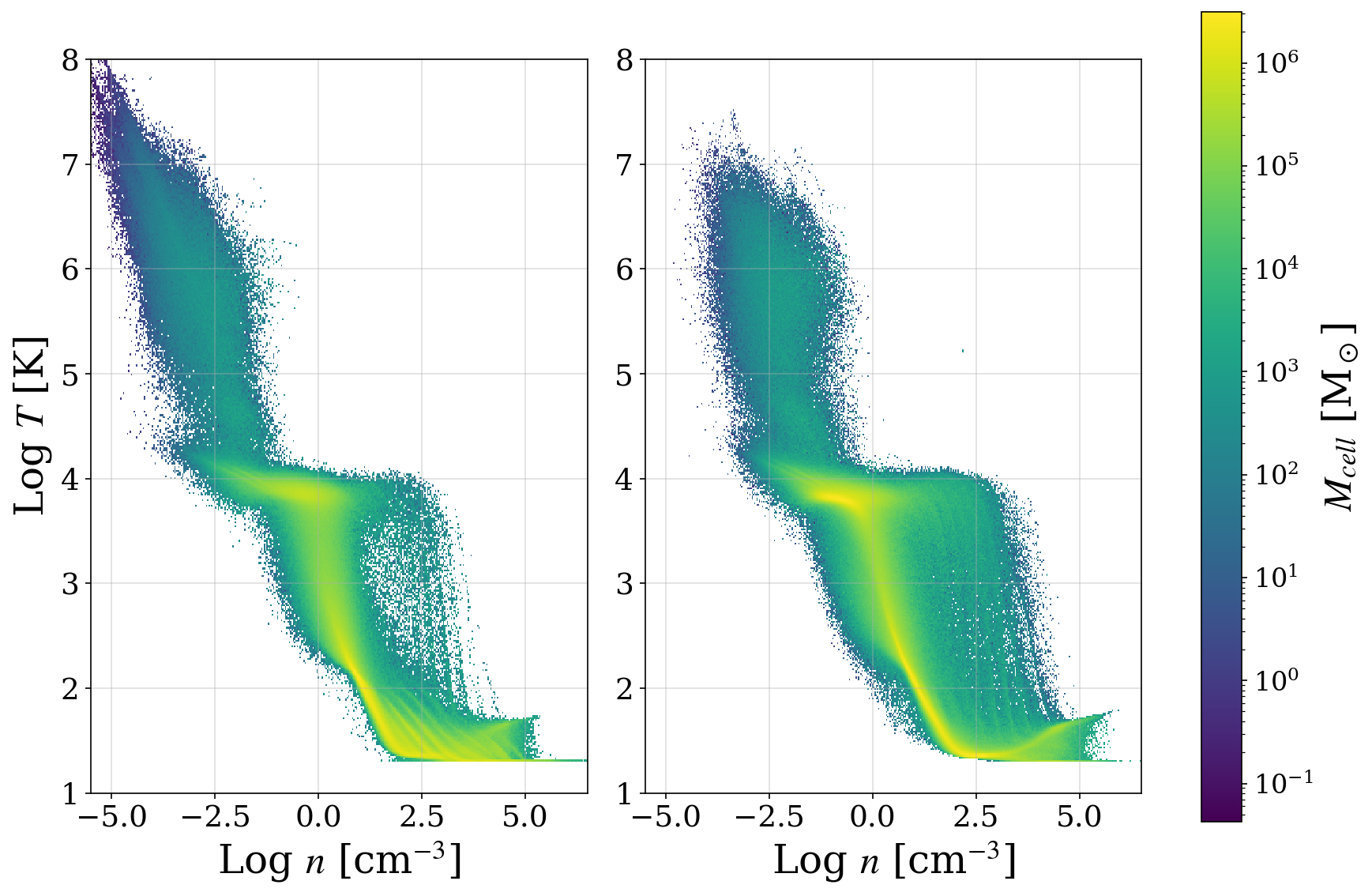}
    \caption[Side-by-side phase plots for the MHD and HD models]{Mass-weighted phase plots of temperature vs.\ total gas number density for the MHD {\em (left)} and HD {\em (right)} simulations. Both simulations are run with a 20\,K temperature floor, hence the sharp cut-off below this temperature.}
    \label{fig:phase}
\end{figure}

The magnetic energy grows exponentially as predicted by dynamo theory. The initial fast growth during the first 75\,Myr is due to the small-scale dynamo. This amplification occurs primarily during the initialisation phase, while the simulation is still responding to the initial conditions. After the first 75\,Myr, the exponential growth of the field energy brings the magnetic energy density to within an order of magnitude of the kinetic energy density. However, our result is consistent with \cite{Pakmor2017MagneticProject}, who find their magnetic energy density saturates at around 10\% of the kinetic energy density. The continuing slow exponential growth after 75\,Myr is consistent with a large-scale dynamo \citep{Gent2024TransitionMedium} if the mean field is being amplified, but could also represent non-linear growth of the small-scale dynamo \citep{Gent2021Small-scaleTurbulence}. The slow growth of the kinetic energy density, perhaps associated with the steady mass loss from the disk, may be a further factor in the increasing magnetic energy density.

At the time of our fiducial snapshot at 150\,Myr, the average volume-weighted field strength is of order 1\,$\upmu$G, lower than the average field strength found for the Milky Way of $\sim 6\,\upmu$G \citep{Beck2001GalacticFields, Pattle2022MagneticCores}, and than observational estimates for the average magnetic field strength in spiral galaxies \citep{Beck2015MagneticGalaxies}. As such, any effect of the magnetic field may be an underestimate. We do not expect our average field strength to match this value however, as our field is not yet saturated, and we do not expect this would saturate until well past $\sim$ 1\,Gyr \citep{Brandenburg2005AstrophysicalTheory}. Due to computational resource constraints and gas depletion via star formation, we can not extend our simulations to these timescales. 


It is important to note that when studying these effects through numerical simulations, the timescale of amplification is strongly dependent on the resolution \citep{Federrath2011ATURBULENCE, Rieder2016APhase, Gent2021Small-scaleTurbulence,Gent2024TransitionMedium}, with higher resolution resulting in more rapid amplification. Fully resolving the small-scale and large-scale dynamo is not currently feasible in galaxy simulations, such as those presented in this work. Our work does not focus on the amplification rate of the dynamo, as we do not have the necessary resolution or simulation time for such a study. Instead, we primarily aim to generate a magnetic field structure that is self-consistent, rather than explicitly prescribed, with reasonable magnetic field strengths.

\begin{figure}
    \centering
	\includegraphics[width=0.35\textwidth]{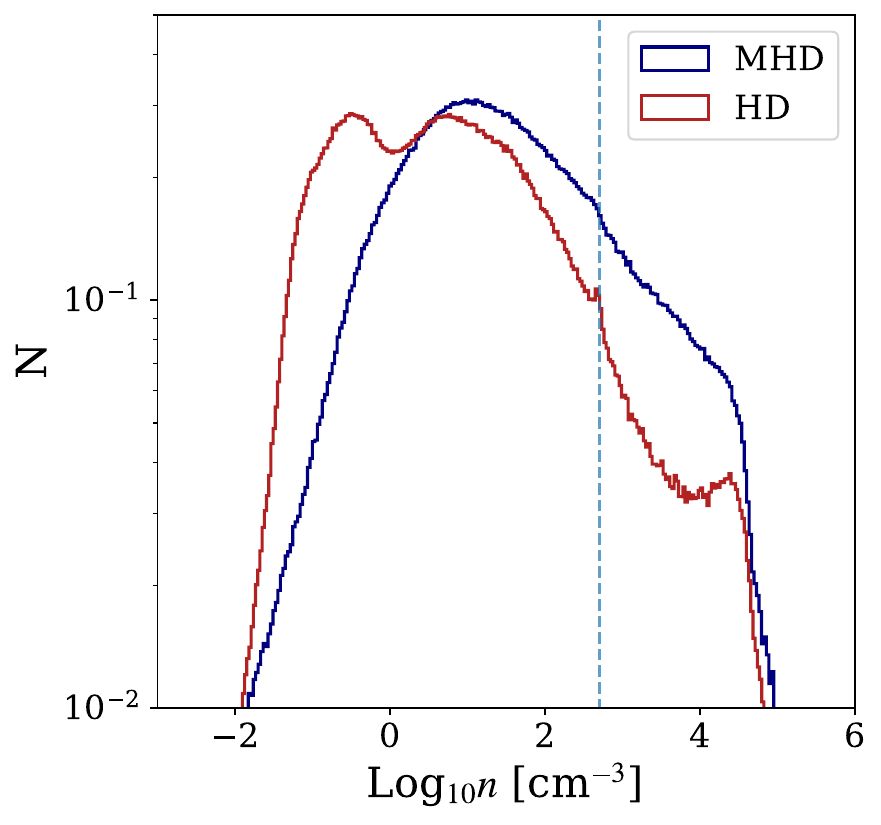}
    \caption[]{Gas density PDFs at $\sim 150$\,Myr for both the MHD {\em (blue)} and HD {\em (red)} simulations. The dashed vertical line indicates the threshold number density above which sink particles can form in bound gas.}
    \label{fig:dens_pdf}
\end{figure}

\subsection{Gas Phase}
\label{subsec:gas_phase}

We examine the density and temperature distribution of the ISM in the disc of both galaxies in Figure~\ref{fig:phase}. Both simulations form a self-consistent ISM with a hot ionised phase, a warm ionised and neutral phase, and a cold neutral phase. As we utilise a time-dependent, non-equilibrium chemical network, we also find gas in the thermally unstable transitions between the different phases. The temperature floor is visible as a flat cut-off at 20K. As in all previous analysis, we consider only gas within the disc. Comparing the two models, we see the range of densities for gas at the same temperature is higher in the HD simulations. 

The probability density functions (PDFs) for gas density in the MHD and HD simulations at $\sim 150$\,Myr are shown in Figure~\ref{fig:dens_pdf}. The dashed line denotes the threshold number density above which sink particle checks are active ($\sim 850\mbox{ cm}^{-3}$). Above this threshold, gravitationally bound gas is missing from the distribution as it is inside sink particles. In this regime, it is difficult to compare the MHD and HD distributions as they are affected in different ways by sink formation. At number densities below the threshold, we still see clearly different distributions, with two peaks in the HD case, and only a single peak in the MHD case. In the MHD case, higher gas densities are more likely than in the HD case---the median density (by mass) in the MHD case is $\sim 22.7\mbox{ cm}^{-3}$, as compared to $\sim 4.6\mbox{ cm}^{-3}$ for the HD case. 

The MHD disc has a higher proportion of dense gas that is not unambiguously bound (i.e., not within sink particles). The characteristic power-law behavior of the PDF at high densities shows a shallower slope, again suggesting that high-density gas is less bound \citep{Klessen2000OnePoint,Kainulainen2009ProbingEvolution,Chen2018TheAnatomy}.

\begin{figure*}
    \centering
	\includegraphics[width=0.7\textwidth]{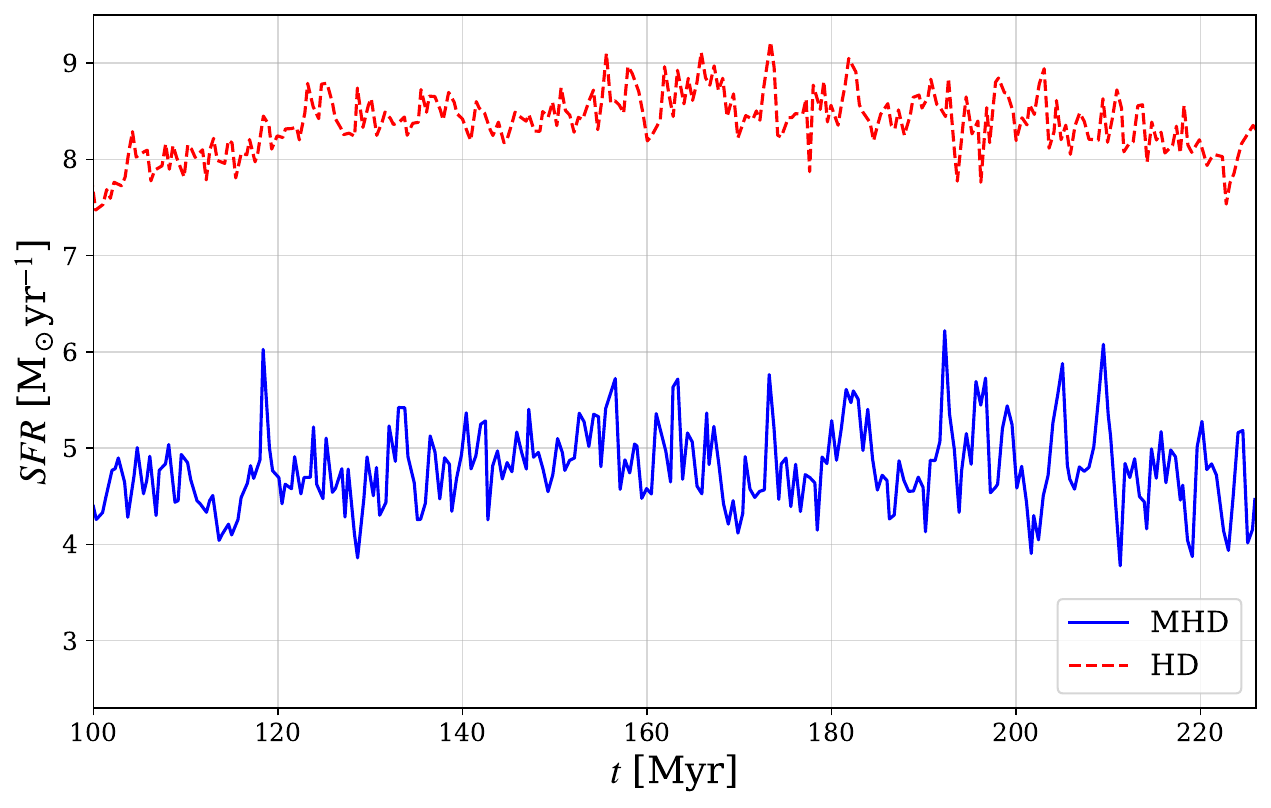}
    \caption[Star formation rates of the MHD and HD simulations]{Star formation rates for both the MHD {\em (solid blue line)} and the HD {\em (dashed red line)} simulations after the initialisation period. The MHD simulation has a consistently lower SFR over time.}
    \label{fig:SFR}
\end{figure*}

\subsection{Star Formation Rate}
\label{subsec:sfr}


In Figure~\ref{fig:SFR}, we show the SFRs for both galaxies across the evolved period. SFRs are calculated for a given snapshot by comparing to the previous snapshot. The stellar mass of all new sink particles (i.e 5\% of the total sink mass), $M_{\odot,\mathrm{new}}$, for the snapshot is added to the stellar mass increase of sink particles present in the previous snapshot, $M_{\odot,\mathrm{acc}}$, and then divided by the elapsed time, $t_2-t_1$, so that
\begin{equation}
    \mathrm{SFR} \;=\; \frac{M_{\odot,\mathrm{acc}} + M_{\odot,\mathrm{new}}}{t_2 - t_1}\ .
    \label{eq: sfr_calc_eq}
\end{equation}
Snapshots are produced roughly every 0.5\,Myr, and an SFR for each snapshot is shown in Figure~\ref{fig:SFR}. We show here only the SFR values after the initialisation phase (i.e.\ beyond 100\,Myr), as the galaxy model requires time to settle before we can analyse the SFR values produced. Averaging from 125--150\,Myr, we find an MHD SFR of $\sim 4.8\,\mathrm{M_{\odot}}$~yr$^{-1}$ and an HD SFR of $\sim 8.4\,\mathrm{M_{\odot}}$~yr$^{-1}$. The average MHD rate is very stable, though with stochastic oscillations due to local variation; we are showing our highest temporal resolution here. The HD rate seems to gently rise, before falling again, likely due to gas depletion of the disc. The trend is clear: the MHD galaxy has a lower SFR than the HD galaxy, by $\sim$ 3.6\,M$_{\odot}$~yr$^{-1}$. This suppression of the SFR is not seen by \cite{Whitworth2022MagneticGalaxies}, but is broadly in line with other spiral galaxy simulations \citep{Robinson2024RegulatingGalaxy, Wibking2023TheGalaxies}. The discrepancy between this suppression and the results of \cite{Whitworth2022MagneticGalaxies} is likely due to the fact that our field strengths are much higher than in their dwarf galaxy models.


\section{Discussion}
\label{sec:discussion} 

\subsection{Kennicutt-Schmidt Relation and Gas Density}
\label{subsec:ks}

\begin{figure*}
    \centering
	\includegraphics[width=1.0\textwidth]{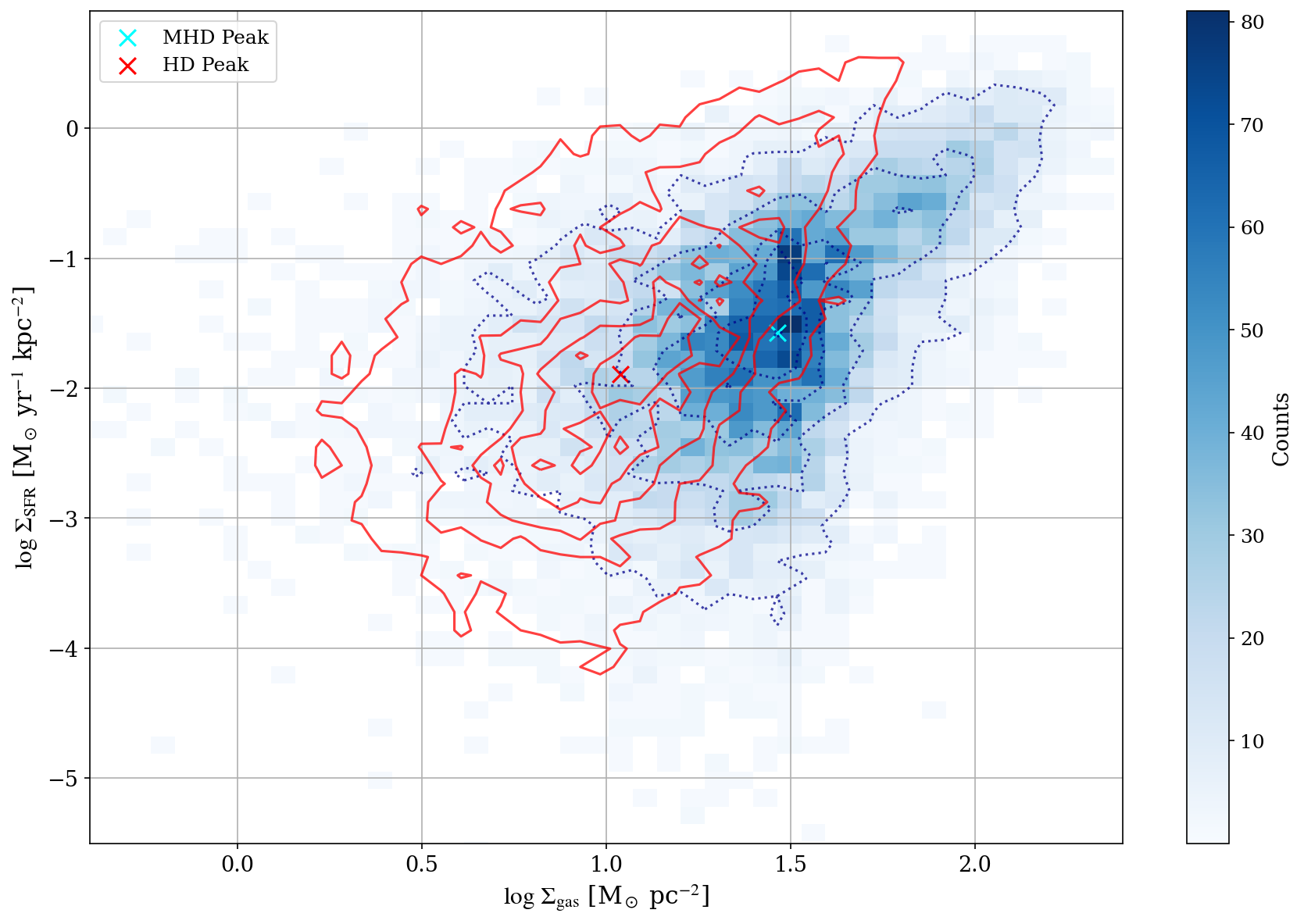}
    \caption[Total gas KS relation for our standard MHD and HD model]{The KS relation for the total gas of both models, measured in $400\times400$\,pc regions, and averaging over the period 125--150\,Myr. The MHD data is shown in {\em blue shading} and {\em blue contours}, with {\em red contours} for the HD data. The peak value for each model is indicated with a cross, {\em cyan} for MHD and {\em red} for HD. Both sets of contours correspond to levels of 6, 24, 42, 60, and finally 78 counts per bin.}
    \label{fig:KS_relation}
\end{figure*}

To further investigate the star formation in our two models, we examine the KS relation. We calculate the total gas surface density by summing the total gas mass in 400\,pc $\times$ 400\,pc regions covering the galaxy disc from -10\,kpc to 10\,kpc in the $x$- and $y$-axes (a total of 2500 regions). We find using 400\,pc $\times$ 400\,pc regions is a good compromise in terms of size-scale. Smaller regions increase the scatter of the relation significantly due to local variation at very small scales, while much larger regions smooth out too much  spatial variation and lead to an extremely narrow range of surface densities that is unrepresentative. For the SFR surface density, we similarly bin all SFR values in the same size regions. We do this for all snapshots between 125 and 150\,Myr, to get a larger statistical sample. The result is shown in Figure~\ref{fig:KS_relation}.

Clearly, the two distributions differ, with the MHD data shifted to higher gas surface densities at the same SFR. The MHD model has a peak at higher gas surface density than the HD, and at a slightly higher SFR surface density. The peak of the KS distribution for the MHD model is at (28.8\,M$_\odot$ pc$^{-2}$, 0.0263\,M$_\odot$ yr$^{-1}$ kpc$^{-2}$), as compared to (11.0\,M$_\odot$ pc$^{-2}$, 0.0129\,M$_\odot$ yr$^{-1}$ kpc$^{-2}$) in the HD case. 


Test models with increased sink resolution were run over a shorter time period, to check the robustness of this KS result, and the shift remains. This is discussed in Appendix~\ref{sec:appendix}. Due to the narrow range of gas surface densities in our models and the large scatter, we do not fit the slope of the relations. Any global trend would likely be determined by the scatter \citep{Heiderman2010THESTUDIES, Whitworth2021IsMetallicities}, and as such we can not confidently make conclusions about any difference in the slope between the two models,

The shift in the relation agrees with Figure~\ref{fig:dens_pdf}, which shows consistently higher gas densities in the MHD case. Our physical interpretation is that this is evidence of additional support by the magnetic field, which makes it more difficult for gravity to become dominant \citep{Kortgen2019GlobalGalaxies, Pattle2022MagneticCores, Wibking2023TheGalaxies, Robinson2024RegulatingGalaxy, Dobbs20232aScales}. Therefore, a greater gas surface density is required to overcome the additional support, begin gravitational collapse, and form stars. 

The difference in the SFR surface density at the peak could be due to the threshold density required for star formation being harder to reach. Therefore, once a region does become bound, it is more massive and able to form stars at a higher rate. On the other hand, regions that would have collapsed and formed stars if there was no magnetic field, cannot collapse in the presence of fields. This means there is no tail of star formation at low surface densities as was seen in the HD case. This could also explain the much more even distribution of star formation in the HD case: as the gas density required for regions to become gravitationally bound is lower in the absence of magnetic fields, more regions are able to form stars. 

\begin{figure}
    \centering
	\includegraphics[width=0.5\textwidth]{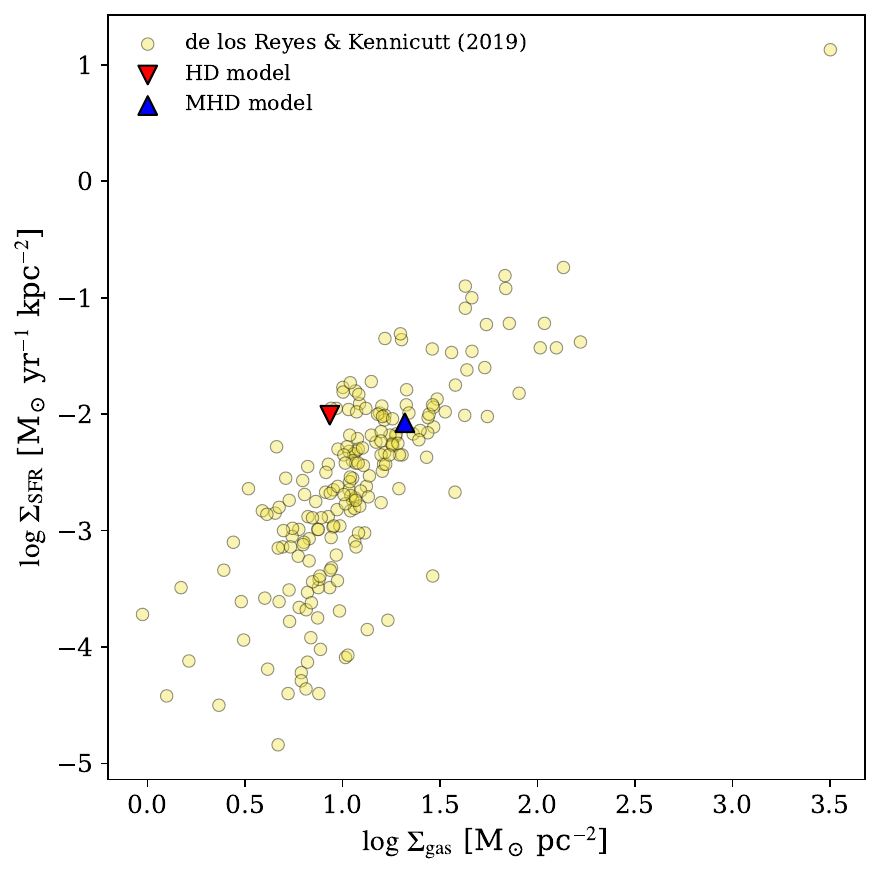}
    \caption[]{Global SFR surface densities and gas surface densities for the MHD and HD model compared to data from \cite{Reyes2019RevisitingGalaxies}.}
    \label{fig:obs_comp}
\end{figure}

In order to compare these two models to observational data, we estimate the average integrated gas surface density and SFR surface density for both models over the 125--150\,Myr time period. As the two models vary in radial extent, we use the radius enclosing 75\% of the gas mass as the radius of disc, and use this to compute the area, and the surface density for each timestep. We calculate the SFR surface density as the SFR divided by this same area, with the SFR computed following the method described in Section \ref{subsec:sfr}. We show the average of these integrated values are in Figure~\ref{fig:obs_comp}, with observational data from \cite{Reyes2019RevisitingGalaxies} for comparison. Both the MHD and HD simulation have similar global SFR surface densities, however the more compact MHD disc has a higher total gas surface density. This puts our MHD model in closer agreement with the observational data. 

\subsection{Caveats}
\label{subsec:caveats} 

While we find that a reduction in sink particle size does not change our KS result, where the MHD galaxy is shifted to higher gas surface densities, this could have a wider impact on other results, such as the magnetic field amplification. We also note the lack of photoionisation and early feedback, which have significant effects on the ISM and the SFR. This is mitigated slightly by our clustered SNe, as the first SN clears out the surrounding area for subsequent SNe as early feedback would. This is purely due to computational constraints. We also shut off accretion onto sink particles after 2\,Myr in order to approximate the effects of early feedback, but this is still only an approximation. 

The lack of early feedback, combined with the isolated nature of the simulated galaxies, and thus lack of gas replenishment, results in our galaxies becoming quite gas poor relatively quickly, compared to a standard star-forming main-sequence galaxy. This limits the total length of time we can evolve our simulations. Simulating for a longer time would allow for a greater amplification of the field by the large-scale dynamo, so our field strengths are lower than we would find at a later time. 

We have also neglected cosmic rays, due to computational constraints. We expect they would lead to a thicker disc and greater gas outflow from the galactic midplane \citep{Girichidis2016LAUNCHINGMEDIUM, Girichidis2018CoolerOutflows}. 

\section{Conclusions}
\label{sec:conclusions} 

In this work, we have presented two isolated galaxy simulations, one with and one without magnetic fields. Our use of sink particles to model star formation ensures that stars only form when gas is unambiguously gravitationally bound. In this way we do not assume or force a KS relation through our algorithm. We also generate a self-consistent magnetic field through dynamo action rather than assuming any particular magnetic field morphology or strength. In our MHD simulation, we start from a small initial seed field, which amplifies to an average volume-weighted field strength of $\sim$ 2.5\,$\upmu$G by the end of our simulation, due primarily to small-scale dynamo action driven by SN feedback (the slower acting large-scale dynamo will not have a large effect over the short simulation timescale). This is only a factor of two smaller than the average field strength observed in the Milky Way, with a field structure consistent with the dynamics of the gas. At the time of our 150\,Myr fiducial snapshot, the average field strength is of order 1\,$\upmu$G. We use a time-dependent chemical network, as well as including radiative heating and cooling. The ISM phase structure includes the hot ionised, warm ionised, warm neutral, and cold neutral phases, as well as thermally unstable gas. We ran both our MHD and HD models for a total of $\sim$ 225\,Myr and find distinct differences between the two.


\begin{itemize}
    \item The MHD model displays a less uniform  gas distribution, with higher peak column densities in more confined regions. The HD galaxy displays more bubbly sub-structure, and its star formation is more evenly distributed than in the MHD case.
    \item The MHD model is more radially compact than the HD equivalent, with a sharper decline in the radial profile. By mass, 75\% of the gas is contained in a 5.1\,kpc radius for the MHD model, compared to 7.4\,kpc for the HD.. 
    \item The MHD model has an extended diffuse atomic envelope above and below the disc, which is not seen in the HD model. We suspect this is due to the rapid generation of magnetic fields in the poloidal direction, resulting in diffuse gas flowing to greater heights above and below the disc.
    \item The MHD model has a lower SFR than the HD model, with an average of $4.8\,\mathrm{M_{\odot}}$~yr$^{-1}$ from 125--150\,Myr, compared to $ 8.4\,\mathrm{M_{\odot}}$~yr$^{-1}$ in the HD case. 
    \item We find a shift in the resolved KS relation to higher gas surface densities in the MHD case. 
    \item We estimate the integrated gas surface density and SFR surface density for both models and compare to observational data from \cite{Reyes2019RevisitingGalaxies} in Figure~\ref{fig:obs_comp}. While both simulations have similar global SFR surface densities, the more compact MHD disc has a higher total gas surface density, putting it in closer agreement with the observational data. 
    \item We interpret the higher amounts of dense, gravitationally unbound gas and lower SFR of the MHD model as arising from additional support from the magnetic field against gravitational collapse. This results in a higher threshold density for star formation in the presence of magnetic fields and thus shifts the KS relation to requiring higher gas surface density $\Sigma_{\mathrm{gas}}$ for a given SFR surface density $\Sigma_{\mathrm{SFR}}$.
\end{itemize}


\section*{Acknowledgements}
We thank R. Pakmor for vital support of the MHD version of {\sc Arepo}. 
This work used the DiRAC@Durham (COSMA) facility managed by the Institute for Computational Cosmology on behalf of the STFC DiRAC HPC Facility (www.dirac.ac.uk). The equipment was funded by BEIS capital funding via STFC capital grants ST/P002293/1, ST/R002371/1 and ST/S002502/1, Durham University and STFC operations grant ST/R000832/1. DiRAC is part of the National e-Infrastructure.
KRJB gratefully acknowledges Peckham Library for providing a workspace that facilitated the production of this manuscript.
M-MML acknowledges support from US NSF grant AST23-07950 and thanks the Inst.\ f\"ur Theoretische Astrophysik for hospitality. 
The team in Heidelberg acknowledges financial support from the European Research Council via the ERC Synergy Grant ``ECOGAL'' (project ID 855130),  from the German Excellence Strategy via the Heidelberg Cluster of Excellence (EXC 2181 - 390900948) ``STRUCTURES'', and from the German Ministry for Economic Affairs and Climate Action in project ``MAINN'' (funding ID 50OO2206). The team in Heidelberg is grateful for computing resources provided by the Ministry of Science, Research and the Arts (MWK) of the State of Baden-W\"{u}rttemberg through bwHPC and the German Science Foundation (DFG) through grants INST 35/1134-1 FUGG and 35/1597-1 FUGG, and also for data storage at SDS@hd funded through grants INST 35/1314-1 FUGG and INST 35/1503-1 FUGG. RSK also thanks the Harvard-Smithsonian Center for Astrophysics and the Radcliffe Institute for Advanced Studies for their hospitality during his sabbatical, and the 2024/25 Class of Radcliffe Fellows for highly interesting and stimulating discussions. 
NB and JG gratefully acknowledge the scientific support and HPC resources provided by the Erlangen National High Performance Computing Center (NHR@FAU) of the Friedrich-Alexander-Universität Erlangen-Nürnberg (FAU) under the NHR project a104bc. NHR funding is provided by federal and Bavarian state authorities. NHR@FAU hardware is partially funded by the German Research Foundation (DFG) – 440719683.
NB further acknowledges support from the ANR BRIDGES grant (ANR-23-CE31-0005). JG is a fellow of the International Max Planck Research School for Astronomy and Cosmic Physics at the University of Heidelberg (IMPRS-HD).

\section*{Data Availability}

Column density projection data, absolute magnetic field projection data, and stacked Kennitcutt-Schmidt data is available via Zenodo (DOI 10.5281/zenodo.17641627). Snapshot data, and assistance in using it, is available on request.



\bibliographystyle{mnras}
\bibliography{references} 



\appendix

\section{Robustness Tests} 
\label{sec:appendix} 

\begin{figure}
    \centering
	\includegraphics[width=0.5\textwidth]{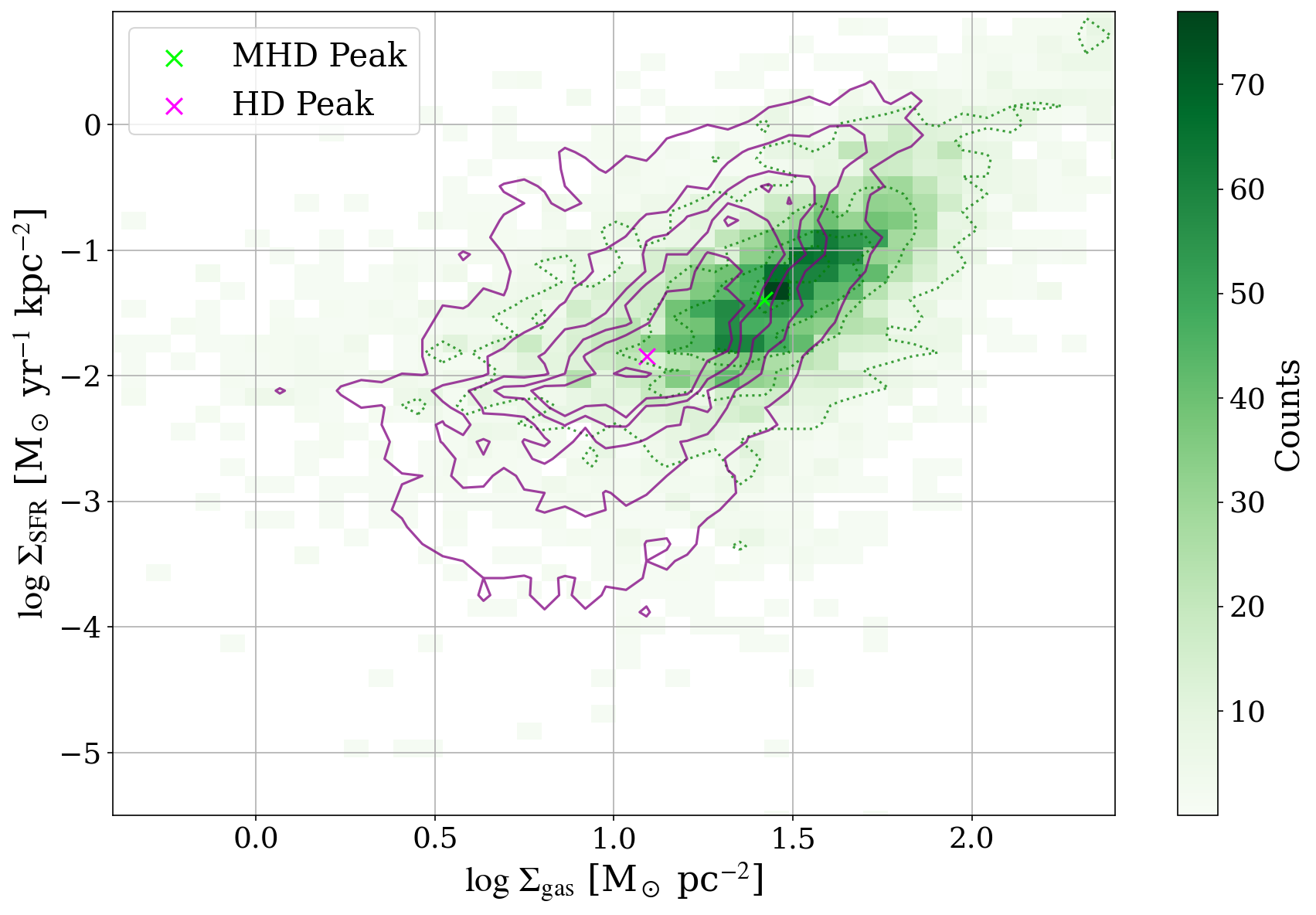}
    \caption[]{Total gas KS Relation for the MHD and HD models with sink particles accretion radius of 2 pc rather than the default 5 pc. MHD data is shown in {\em green} shading and contours, HD in {\em purple} contours. We see that the shift to higher gas densities in the MHD case remains. As in the original figure, both sets of contours correspond to levels of 6, 24, 42, 60, and finally 78 counts per bin.}
    \label{fig:KS_relation_small_sinks}
\end{figure}

In this paper we analyse the KS relation, binning our data at a spatial scale of 400\,pc. We identify the presence of dense, extended clumps in our MHD simulation, which can be a sign that the simulation is not sufficiently resolved. This leads to the question of whether or not these clumps are physical, and if they are large enough to affect our KS analysis. To test this we would ideally use a third simulation that does not display such clumps, on which we can perform the same analysis to see if the result is robust.

Visually similar clumpy structures have been noted in previous HD simulations, attributed to a lack of gas cell resolution \citep{Truelove1997TheHydrodynamics} and violation of the Truelove criteria \citep[see for example][who explicitly show this in their Figure 1]{Greif2011SimulationsProtostars}. In HD simulations, these artefacts dissipate when resolution of the gas is increased, as we ourselves found in early tests of our own HD models. However, the same increase in gas resolution in our MHD model did not lead to the dissipation of these artefacts. Instead, we found that reducing the size of sink particles did produce a notable difference. The mechanism that explains this is not certain, and would require further investigation. It is possible that as sink particles accrete gas, but not magnetic flux (this would violate energy and flux conservation), that magnetic flux is ``left behind" and builds up in the vicinity of sink particles. This would lead to an unphysical increase in the magnetic field strength, perhaps supporting gas against collapse. It is also possible that the collapse of gas around sinks in creating an over-exaggerated small-scale dynamo effect, exasperated by a lack of suitable resolution to resolve magnetic field effects at the sink accretion radius. Irrespective of the origin, using a smaller sink particle size can allow us to test the robustness of our KS result on a model without such extended dense clumps.

However, simulations with a reduced sink size is more computationally expensive, which prevents us from completing a full simulation with this choice. As an alternative, we restart our standard MHD model at $\sim$ 100\,Myr, but using sink particles with a formation and accretion radius reduced from 5\,pc to 2\,pc. We run for a further $\sim$ 20\,Myr, twice the typical lifetime of a molecular cloud \citep{Chevance2022TheClouds}, which should allow changes in cloud properties to fully develop. We show the result of this new KS analysis, with data stacked from 100 to 120\,Myr, in Figure~\ref{fig:KS_relation_small_sinks}, with MHD data in green shading and green contours and HD data in purple contours. We find an almost identical shift between the two models, assuring us that the result is robust to changes in sink particle size, and as such is not greatly affected by the extended clumps. Naturally we would expect more resolved analysis (on scales smaller than 400\,pc) to be more affected by these structures.

\section{Gas Type} 
\label{sec:appendix_chem}

\begin{figure*}
    \centering
    \includegraphics[width=0.995\textwidth]{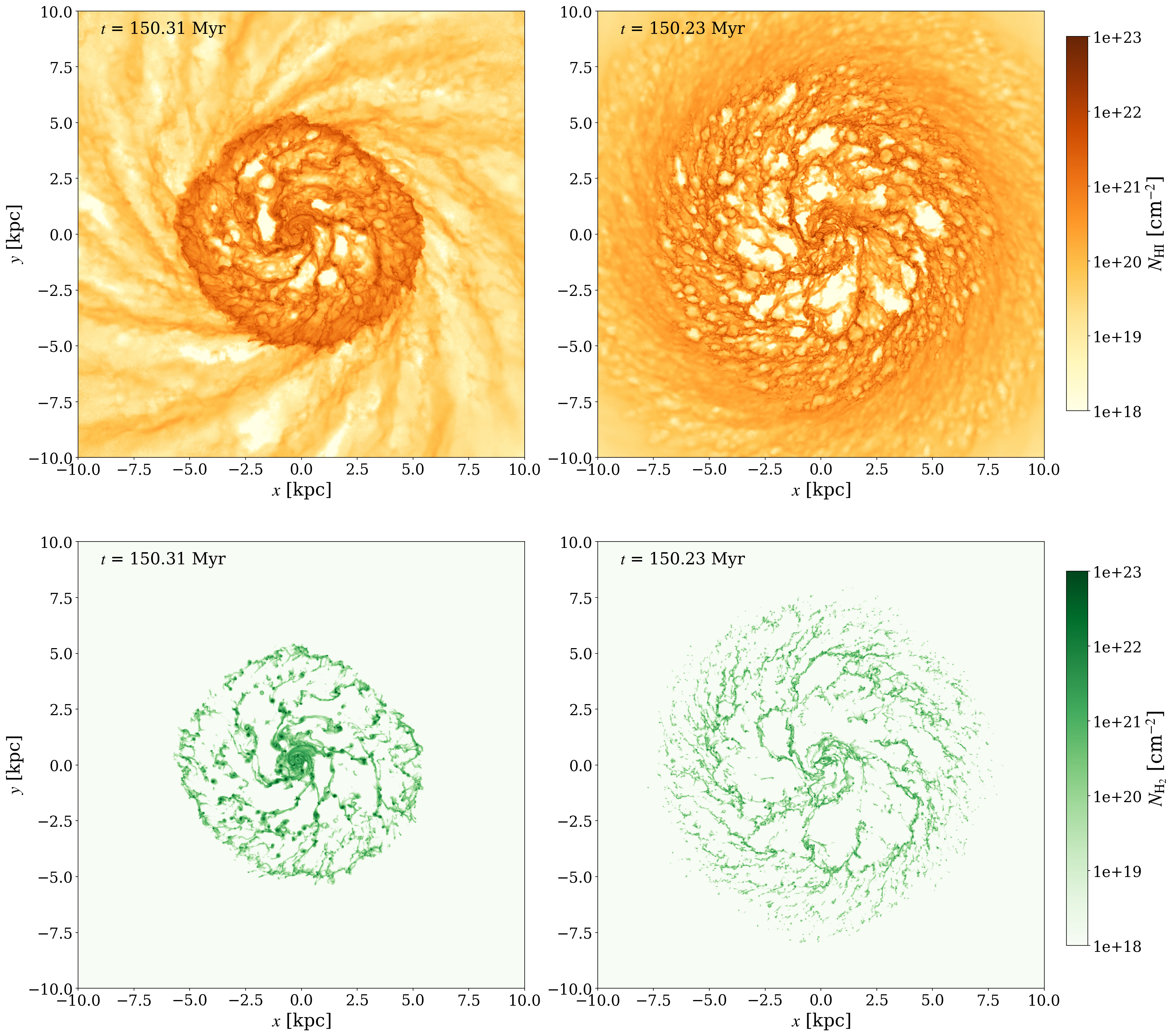}
    \caption{Column densities of H\,\textsc{i} {\em (top panels)} and H$_2$ gas {\em (bottom panels)} for both the MHD (left) and HD (right) models at 150\,Myr.
    \label{fig:HI_H2}}
\end{figure*}

The chemical network included in these models also allows us to trace the evolution of  H\,\textsc{i} and H$_2$ gas. The distribution of H\,\textsc{i} and H$_2$ gas (Figure~\ref{fig:HI_H2}) shows similar trends to the total gas (Figure~\ref{fig:proj_with_sinks}) with the MHD model being more compact and having higher peak column densities, while the HD model has a more uniform distribution. 


\bsp	
\label{lastpage}
\end{document}